\documentclass[aps,twocolumn,10pt,superscriptaddress,showkeys,groupedaddress]{revtex4-1}
\usepackage{amssymb}
\usepackage{epsfig} 
\usepackage{amsmath} 
\usepackage{graphicx}
\usepackage{color}
\usepackage{float}
\usepackage{xcolor}
\usepackage{amssymb}
\usepackage{enumitem}
\usepackage{natbib}

\definecolor{red}{rgb}{0.8,0,0}
\definecolor{violet}{rgb}{0.4,0,0.4}
\definecolor{green}{rgb}{0,0.5,0.0}
\definecolor{navy}{rgb}{0.0,0.0,0.6}
\definecolor{orange}{rgb}{0.8,0.2,0.0}

\usepackage[normalem]{ulem}  

\newcommand{\physrep}{Phys. Rep.}
\newcommand{\apjl}{Astro. Phys. J. Lett.}

\newcommand{\aap}{Astronomy $\&$ Astrophysics}
\newcommand{\nphysa}{Nuc. Phys. A}

\begin{document} 

\title{Dense matter equation of state of a massive neutron star with anti-kaon condensation}

\author{Vivek Baruah Thapa}
\email{thapa.1@iitj.ac.in}

\author{Monika Sinha}
\email{ms@iitj.ac.in}

\affiliation{Indian Institute of Technology Jodhpur, Jodhpur 342037, India} 
\date{\today}
\begin{abstract}
{Recent measurements of neutron star mass from several candidates (PSR J$1614-2230$, PSR J$0348+0432$, MSP J$0740+6620$) set the lower bound on the maximum possible mass for this class of compact objects $\sim 2$ M$_\odot$. Existence of stars with high mass brings the possibility of existence of exotic matter (hyperons, meson condensates) at the core region of the objects. In this work, we investigate the (anti)kaon ($K^-, \bar{K}^0$) condensation in $\beta-$equilibrated nuclear matter within the framework of covariant density functional theory. The functionals in the kaonic sector are constrained by the experimental studies on $K^-$ atomic, kaon-nucleon scattering data fits. We find that the equation of state softens with the inclusion of (anti)kaon condensates, which lowers the maximum mass of neutron star. In one of the density-independent coupling cases, the $K^-$ condensation is through a first-order phase transition type, which produces a $2$ M$_\odot$ neutron star. The first-order phase transition results in mixed phase region in the inner core of the stars. While $\bar{K}^0$ condensation appears via second-order phase transition for all the models we consider here.}
\end{abstract}
\maketitle

\section{Introduction}
Being born as a consequence of supernova explosion, the neutron stars provide an idiosyncratic environment to study the exotic phases of matter at high densities which may range from a few to several times nuclear saturation density \cite{1996cost.book.....G, 2007PrPNP..59...94W}. Such high density ranges could never be achieved in any terrestrial laboratories now or in the near future at low temperature as present inside these compact objects. Consequently till now the true nature of matter above nuclear saturation density is a matter of speculation. The neutron star being very compact is subject to tremendous inward gravitational pull which is supported by the degeneracy pressure mainly of neutrons with small admixture of other baryon and lepton species. The exact nature of inter-particle interaction at such high densities is not known precisely, though many authors have provided various phenomenological models based on density functional theories \cite{1972PhRvC...5..626V,1998NuPhA.637..435S,2001A&A...380..151D,2008PhRvC..78f5805S, 2014PhRvC..89d5807B} and realistic nuclear potentials \cite{1998PhRvC..58.1804A,2006PhRvC..74d7304L, 2015RvMP...87.1067C,2017PhRvC..96c4307H,2019PhRvC.100d5803L} to explain and understand the nature of highly dense matter and internal structure of these compact stars. Relativistic-Mean Field (RMF) theory, also known as Walecka model is one of the foremost models \cite{1974AnPhy..83..491W}. This model produces the saturation properties of nuclear matter and finite nuclei to a good extent. However, in order to explain the saturation properties viz. compression modulus and effective mass, non-linear self-interaction terms for the scalar meson fields has been incorporated in this model \cite{Walecka1986,1996PhRvC..53.1416S}. In addition, other self-interacting vector meson fields are also brought into consideration to explain the interaction in high density regimes \cite{1991NuPhA.526..703B, 1992ZPhyA.342..387G, 1996NuPhA.598..539F}. But at those regimes due to the dependence on higher orders of fields, instabilities may arise. The density-dependence of the meson-baryon couplings (DDRH model) is another possible approach to reduce the instabilities as in the later case the higher orders field terms are not brought into consideration \cite{2005PhRvC..71f4301T,2010PhRvC..81a5803T}. For thermodynamic consistency, a re-arrangement term is considered which contributes explicitly to the matter pressure consequently influencing the equation of state at higher densities.

Nuclear matter is composed of mainly neutrons with small admixture of protons and electrons in $\beta$-equilibrium condition and fraction of protons are electrons are equal to keep the condition of charge neutrality. With the increase of neutron density the electron density and hence the Fermi momentum increases to keep the matter in $\beta$-equilibrium. When the Fermi momentum of electron reaches the mass of muons, muons appear. With further increase of density when the electron Fermi energy reaches the vacuum mass of mesons (pion or kaon), condensate of negatively charged mesons starts to appear which in turn help to maintain the charge neutrality. However, $s$-wave $\pi N$ scattering potential being repulsive the effective ground state mass of $\pi$ meson increases \citep{1985ApJ...293..470G} opposing the possibility of $\pi$-meson appearing. But, effective ground state mass of $K$-meson decreases due to its attractive interaction with nucleons opening the possibility of $K$-meson appearing. Kaplan and Nelson \cite{1988NuPhA.479..273K,1987PhLB..192..193N} for the very first time demonstrated that antikaon $K^-$ may undergo Bose-Einstein condensation in dense matter formed in heavy-ion collisions. Furthermore, other evidences such as the $K^-$ atomic data, kaon-nucleon scattering data \cite{1995NuPhA.585..401L,1994NuPhA.567..937B,1994PhLB..326...14L} studied by several authors in chiral perturbation theory also encouraged the concept of $K^-$ condensed phase presence in the interior of neutron star. The in-medium energy of (anti)kaon $K^-$ mesons decreases in the dense matter due to the lowering of effective mass. Finally, the onset of $s$-wave $K^-$ condensation occurs when the chemical potential of $K^-$ ($\omega_{K^-}$) equates the electron chemical potential ($\mu_{e}$). The $s$-wave $\bar{K}^0$ appears when its chemical potential ($\omega_{\bar{K}^0}$) equates to zero. The threshold density of (anti)kaon appearance is very sensitive to the optical potential in nuclear symmetric matter. Studies \cite{1994NuPhA.567..937B,1994PhLB..326...14L, 1995PhRvC..52.3470K, 1996PhRvC..53.1416S} reveal that $K^+$ mesons develop a repulsive optical potential nature in the nuclear matter. Thus, it may be concluded that kaon $K^+$ condensation is not favored inside the neutron star. Phase transitions from hadronic to kaonic phases in dense matter may be either a first (mixed phase) or second order type depending on the (anti)kaon optical potential depths. Various observational features of neutron star evolution such as the spin down rates, cooling, glitches may be affected by the alterations of weak interaction rates, transport properties of matter interior to neutron star due to phase transitions \cite{2000PhR...328..237H,2003NuPhA.720..189K}. The first order phase transition cannot be explained by merely the Maxwell's construction accounting for only one charge conservation because neutron stars have two conserved charges viz. baryon number conservation and global charge neutrality. The Gibbs conditions are employed to adequately explain the mixed phase regime of the neutron star interior \cite{1996cost.book.....G}.

The composition and equation of state (EOS) of matter inside the compact stars are constrained by recent observations of compact objects in wide range of electromagnetic spectra and in gravitational wave. Most fundamental property of compact objects which constrain the EOS is the observed mass of neutron stars (NSs) which sets the limit of maximum possible mass of NS family. One of the most accurately measured pulsar masses was done by Hulse and Taylor of binary pulsar PSR $1913+16$ with mass $1.4408 \pm 0.0003$ M$_\odot$ \cite{1975ApJ...195L..51H}. In recent years, several new neutron star mass measurements $> 2$ M$_\odot$ were accomplished viz. the millisecond pulsar (MSP) PSR J$1614-2230$ of mass $1.97 \pm 0.04$ M$_\odot$ \cite{2010Natur.467.1081D,2010ApJ...724L.199O}, PSR J$0348+0432$ ($2.01\pm 0.04$ M$_\odot$) \cite{2013Sci...340..448A} and MSP J$0740+6620$ ($2.14^{+0.20}_{-0.18}$ M$_\odot$ with 95$\%$ credibility) \cite{2020NatAs...4...72C}. Gravitational wave observations have been also successful in providing bounds on the compact object mass. The recent `GW190814' event observed by the LIGO-Virgo Collaboration (LVC) from a coalescence of a black hole and a lighter companion sets the mass of the former to be $23.2^{+1.1}_{-1.0}$ M$_\odot$ and that of the latter to be $2.59^{+0.08}_{-0.09}$ M$_\odot$ \cite{2020ApJ...896L..44A}. The nature of the lighter companion to be a heavy neutron star or a light black hole is still not clear. Recently, the space mission NICER (Neutron star Interior Composition ExploreR) also provided mass-radius measurements of PSR J$0030+0451$ to be $1.44^{+0.15}_{-0.14}$ M$_\odot$, $13.02^{+1.24}_{-1.06}$ km \cite{2019ApJ...887L..24M} and $1.34^{+0.15}_{-0.16}$ M$_\odot$, $12.71^{+1.14}_{-1.19}$ km \cite{2019ApJ...887L..21R} respectively.

The existence of massive NSs open the possibility of exotic matter appearance in the inner core of the star. The addition of exotic degrees of freedom enhances the softening of equation of states (EOSs). But with the observation of neutron stars possessing mass greater than $2$ M$_\odot$, the softer EOSs may be mostly discarded as they lead to stars with lower maximum mass \cite{1996cost.book.....G}. Various non-linear density-independent parametrizations viz. GM1, GM2, GM3 \cite{1991PhRvL..67.2414G}; TM1 \cite{1994NuPhA.579..557S} evaluates relatively softer EOSs which cannot develop $2$ M$_\odot$ neutron stars with (anti)kaon condensation \cite{2001PhRvC..63c5802B,2000NuPhA.674..553P} undergoing $1^{\text{st}}-$order phase transition. GMT parametrization \cite{2000NuPhA.674..553P} generates a relatively stiffer EOS thus producing a mixed phase region in the neutron star interior. The GMT set is generated by reproducing the saturation properties of TM1 parametrization with the exclusion of non-linear self-interaction $\omega$-meson (vector) term exclusion. Density-dependent model parametrizations such as DD2 are able to produce stiff EOSs with baryon octet and (anti)kaon condensates as composition leading to neutron stars possessing mass $\geq$ $2$ M$_\odot$ \cite{2014PhRvC..90a5801C}.

In this work, we explore the possibility of (anti)kaon condensation in $\beta$-equilibrated nuclear matter in the inner core within the non-linear and density-dependent covariant density functional (CDF) model which shows consistent results with recent astrophysical observations.

The paper is organized in the following manner. In Sec.~\ref{sec:formalism} we briefly discuss the non-linear CDF and density-dependent relativistic hadron (DDRH) field theory formalism incorporated in this work. Our results are shown in Sec.~\ref{sec:results} and our conclusions are summarized in Sec.~\ref{sec:conclusions}.

\section{Formalism} \label{sec:formalism}

\subsection{Model}

In this section, we introduce the non-linear and density dependent CDF model to study the phase transition from hadronic to anti-kaon condensed matter which could be either of the first-order or second-order form. Throughout the work, for the hadronic matter we have considered nucleons ($N\equiv n,p$) alongside electrons and muons. The strong interactions between the baryons as well as the anti-kaons are mediated by the scalar $\sigma$, isoscalar-vector $\omega^\mu$ and isovector-vector $\rho^{\mu \nu}$ meson fields. We have considered the mean field model with non-linear (NL) scalar meson self interaction as well as density-dependent coupling constants. Throughout the model, the implementation of natural units is incorporated ($\hbar=c=1$). In general, the total Lagrangian density of the matter is given by  \cite{1999PhRvC..60b5803G,2000NuPhA.674..553P,2001PhRvC..64e5805B,2001PhRvC..63c5802B,2000NuPhA.674..553P,1999PhRvC..60b5803G}
\begin{equation} \label{eqn.2}
    \begin{aligned}
\mathcal{L} & = \sum_{N} \bar{\psi}_N(i\gamma_{\mu} D^{\mu} - m^{*}_N) \psi_N + \sum_{l} \bar{\psi}_l (i\gamma_{\mu} \partial^{\mu} - m_l)\psi_l \\
 & + D^*_\mu \bar{K} D^\mu K - m^{*^2}_K \bar{K} K + \frac{1}{2}(\partial_{\mu}\sigma\partial^{\mu}\sigma - m_{\sigma}^2 \sigma^2) \\ 
 & -  \frac{1}{4}\omega_{\mu\nu}\omega^{\mu\nu} + \frac{1}{2}m_{\omega}^2\omega_{\mu}\omega^{\mu} - \frac{1}{4}\boldsymbol{\rho}_{\mu\nu} \cdot \boldsymbol{\rho}^{\mu\nu} + \frac{1}{2}m_{\rho}^2\boldsymbol{\rho}_{\mu} \cdot \boldsymbol{\rho}^{\mu} \\
& - \text{U}(\sigma), \quad \quad \text{[Only for NL model}].
    \end{aligned}
\end{equation}
Here, U$(\sigma)$ stands for the self interactions of scalar mesons which is present in case of NL model, but not considered in case of density-dependent couplings, the fields $\psi_N$, $\psi_l$ correspond to the baryon and lepton fields with their bare masses, $m_N$ and $m_l$ respectively. The covariant derivative is given by
\begin{equation}
D_\mu = \partial_\mu + ig_{\omega j} \omega_\mu + ig_{\rho j} \boldsymbol{\tau}_j \cdot \boldsymbol{\rho}_{\mu}    
\end{equation}
with `$j$' denoting the nucleons and (anti)kaons. The isospin doublets for kaons are denoted by $K\equiv (K^+,K^0)$ and that for anti-kaons by, $\bar{K} \equiv (K^-, \bar{K}^0)$. The effective nucleon (Dirac) and anti-kaon masses in the mean-field approximation are given by
\begin{equation} \label{eqn.6}
\begin{aligned}
    m_{N}^* = m_N - g_{\sigma N}\sigma, \quad m_{K}^* = m_K - g_{\sigma K}\sigma
\end{aligned}
\end{equation}
where $m_N$, $m_K$ are the bare nucleon and kaon masses respectively. The field strength tensors for the vector fields in eq.\eqref{eqn.2} are given by
\begin{equation} \label{eqn.4}
\begin{aligned}
\omega_{\mu \nu} & = \partial_{\mu}\omega_{\nu} - \partial_{\mu}\omega_{\nu} ,\\
\boldsymbol{\rho}_{\mu \nu} & = \partial_{\nu}
\boldsymbol{\rho}_{\mu} - \partial_{\mu}\boldsymbol{\rho}_{\nu}
\end{aligned}    
\end{equation}

The scalar self-interaction terms \cite{1977NuPhA.292..413B} required in NL model are given by 
\begin{equation} \label{eqn.3}
    \text{U}(\sigma) = \frac{1}{3} g_2 \sigma^3 + \frac{1}{4} g_3 \sigma^4
\end{equation}
where, $g_2 = b m_N g^3_{\sigma N}$ and $g_3 = c g^4_{\sigma N}$. In the same approximation, the meson fields acquire the ground state expectation values as
\begin{equation} \label{eqn.8}
\begin{aligned}
\sigma & = -\frac{1}{m_{\sigma}^2} \frac{\partial \text{U}}{\partial \sigma} + \sum_{b} \frac{1}{m_{\sigma}^2} g_{\sigma b}n_{b}^s + \sum_{\bar{K}} \frac{1}{m_{\sigma}^2} g_{\sigma K}n_{\bar{K}}^s,\\
  \omega_{0} & = \sum_{b} \frac{1}{m_{\omega}^2} g_{\omega b}n_{b} - \sum_{\bar{K}} \frac{1}{m_{\omega}^2} g_{\omega K}n_{\bar{K}}, \\
    \rho_{03} & = \sum_{b} \frac{1}{m_{\rho}^2} g_{\rho b}
  \boldsymbol{\tau}_{b3}n_{b} + \sum_{\bar{K}} \frac{1}{m_{\rho}^2} g_{\rho K}
  \boldsymbol{\tau}_{\bar{K}3}n_{\bar{K}}
\end{aligned}    
\end{equation}
The first term in right hand side of eq.\eqref{eqn.8} ($\sigma$-meson field) is required only in the case of NL model. In case of DDRH model, the scalar self-interaction terms are absent \cite{2001PhRvC..64c4314H, 2001PhRvC..64b5804H}. The meson-baryons (anti-kaon) couplings are denoted by $g_{ij}$, where $i$ goes over the mesons and $j$ over the baryons and (anti)kaons. $\tau_j$ represents the iso-spin operator. The scalar and baryon(vector) number densities are defined for the baryons as $n_{N}^s= \langle\bar{\psi}_N \psi_N \rangle$, $n_{N}=\langle\bar{\psi}_N \gamma^0 \psi_N\rangle$ respectively. In case of the $s$-wave (anti)kaons condensates, the number density is given by \cite{1999PhRvC..60b5803G}
\begin{equation} \label{eqn.11}
\begin{aligned}
    n_{K^-, \bar{K}^0} & = 2 \left( \omega_{\bar{K}} + g_{\omega K} \omega_0 \pm \frac{1}{2} g_{\rho K} \rho_{03} \right) \\
    & = 2 m^*_K \bar{K} K
\end{aligned}
\end{equation}
The scalar density and vector number density of the baryon-$N$ at zero temperature are given by
\begin{equation} \label{eqn.12}
\begin{aligned}
n^{s}_N & = \frac{m^{*}_{N}}{2 \pi^2} \left[ p_{{F}_{N}} E_{F_N}  - m_{N}^{*^2} \ln \left( \frac{p_{{F}_N} + E_{F_N}}{m_{N}^{*}} \right) \right], \\
n_N & = \frac{p_{{F}_{N}}^{3}}{3 \pi^2},
\end{aligned}    
\end{equation}
respectively, where $p_{{F}_{N}}$ and $E_{F_N}$ are the Fermi momentum and fermi energy of the $N$-th nucleon respectively.
The in-medium energies of $\bar{K}\equiv (K^-,\bar{K}^0)$ for s-wave condensation are provided by
\begin{equation} \label{eqn.7}
    \omega_{K^{-} , \bar{K}^0} = m^*_K - g_{\omega K} \omega_0 \mp \frac{1}{2} g_{\rho K} \rho_{03}
\end{equation}
with the isospin projections for $K^-,\bar{K}^0$ being $-1/2,+1/2$ respectively.

The chemical potential of the $N$-th nucleon is
\begin{equation} \label{eqn.15}
\begin{aligned}
    & \mu_{N} = \sqrt{p_{F_N}^2 + m_{N}^{*2}} + g_{\omega N}\omega_{0} + g_{\rho N} \boldsymbol{\tau}_{N3} \rho_{03} + \Sigma^{r},
\end{aligned}
\end{equation}
where, the rearrangement term $\Sigma^{r}$ is introduced to maintain the thermodynamic consistency in case of DDRH model \cite{2001PhRvC..64b5804H} which is given by
\begin{equation}\label{eqn.22}
\begin{aligned}
\Sigma^{r} & = \sum_{N} \left[ \frac{\partial g_{\omega N}}{\partial n}\omega_{0}n_{N} - \frac{\partial g_{\sigma N}}{\partial n} \sigma n_{N}^s + \frac{\partial g_{\rho N}}{\partial n} \rho_{03} \boldsymbol{\tau}_{N3} n_{N} \right],
\end{aligned}
\end{equation}
where $n= \sum_{N} n_N$ is the total baryon number density. This re-arrangement term contributes  explicitly only to the matter pressure. In case of NL CDF model, this term is not required. 

The total energy density from the baryonic and leptonic matter is given by
\begin{widetext}
\begin{eqnarray} \label{eqn.20}
\begin{aligned}
\varepsilon_f & = \frac{1}{2}m_{\sigma}^2 \sigma^{2} + \frac{1}{2} m_{\omega}^2 \omega_{0}^2 + \frac{1}{2}m_{\rho}^2 \rho_{03}^2 + \sum_N \frac{1}{\pi^2} \left[ p_{{F}_N} E^3_{F_N} - \frac{m_{N}^{*2}}{8} \left( p_{{F}_N} E_{F_N} + m_{N}^{*2} \ln \left( \frac{p_{{F}_N} + E_{F_N}}{m_{N}^{*}} \right) \right) \right] \\
	 & + \frac{1}{\pi^2}\sum_l \left[ p_{{F}_l} E^3_{F_l} - \frac{m_{l}^{2}}{8} \left( p_{{F}_l} E_{F_l} + m_{l}^{2} \ln \left( \frac{p_{{F}_l} + E_{F_l}}{m_{l}} \right) \right) \right]
\end{aligned}
\end{eqnarray}
\end{widetext}

For the anti-kaons condensates, the energy density contribution to the total one is given by
\begin{equation} \label{eqn.21}
    \varepsilon_{\bar{K}} = m^*_K (n_{K^-} + n_{\bar{K}^0})
\end{equation}
Anti-kaons being Bose condensates, there is no direct contribution to the total matter pressure from their ends. The matter pressure is related to the energy density via the thermodynamic relation (Gibbs-Duhem) as
\begin{equation}
    p_m = \sum_{N} \mu_N n_N + \sum_{l} \mu_l n_l - \varepsilon_f,
\end{equation}
as there will be no contribution from the anti-kaons. The total energy density is given as $\varepsilon = \varepsilon_f + \varepsilon_{\bar{K}}$.

Initially at the lower density, matter is composed of neutrons ($n$), protons ($p$) and electrons ($e$) in beta-equilibrium and charge neutrality condition. Beta-equilibrium at this stage is satisfied by the chemical potential balance,
\begin{equation}
    \mu_n = \mu_p + \mu_e
\end{equation}
With increasing density, when the chemical potential of electrons becomes equal to the rest mass of muons, the muons appear. Hence the threshold equilibrium condition for the onset of muons is, $\mu_e = m_\mu$.

Studies \cite{1997PhR...280....1P,1996cost.book.....G} shows that strangeness changing processes such as, $N \rightleftharpoons N + \bar{K}$ and $e^- \rightleftharpoons K^-$ may come into picture inside the neutron star core. Hence kaons may appear by these reactions, when the threshold conditions are satisfied as
\begin{equation} \label{eqn.17}
\begin{aligned}
    \mu_n - \mu_p = \omega_{K^-} = \mu_e, \quad \omega_{\bar{K}^0} = 0
\end{aligned}
\end{equation}

The charge neutrality conditions in the hadronic and kaon condensed phases are given by
\begin{equation} \label{eqn.18}
\begin{aligned}
 Q^h & = \sum_N q_N n^h_N - n_e - n_\mu = 0, \\
 Q^{\bar{K}} & = \sum_N q_N n^{\bar{K}}_N - n_{K^-} - n_e - n_\mu = 0
\end{aligned}
\end{equation}
respectively, where $n^h_N$ and $n^{\bar{K}}_N$ represents the number densities in hadronic and kaon phases respectively, both having the same form.

The appearance of (anti)kaon condensate may occur either through first order or through second order phase transition from the hadronic to kaon phase depending on the optical potential depths of $K^-$ at nuclear saturation density. In case the transition sets through first order form, the mixed phase comes into picture: the two phases of pure hadronic matter without condensate and with condensate co-exist. Then for this case, the Gibbs conditions alongside global baryon number conservation and charge neutrality can be enforced to determine the mixed phase state \cite{1999PhRvC..60b5803G,1998PhRvL..81.4564G,1992PhRvD..46.1274G}. The Gibbs conditions for this state are,
\begin{eqnarray} \label{eqn.23}
    p^{h}_m & = p^{\bar{K}}_m, \\
    \mu^h_N & = \mu^{\bar{K}}_N \label{eqn.24}
\end{eqnarray}
where $h$ and $\bar{K}$ superscripts represent the respective quantities in hadronic and anti-kaon condensed phase respectively. In addition, two additional global constraints (viz. global baryon number conservation and charge neutrality) are enforced via the relations,
\begin{eqnarray} \label{eqn.25}
    n_N & = (1 -\chi) n^h_N + \chi n^{\bar{K}}_N, \\
    (1 & -\chi) Q^h + \chi Q^{\bar{K}} = 0 \label{eqn.26}
\end{eqnarray}
respectively. Here, $\chi$ is the fraction of anti-kaon ($K^-$) condensed phase in the mixed phase regime. $\chi \sim 0, 1$ represents the initiation and termination of mixed phase region respectively. Region with $\chi<0$ is the pure hadronic phase and $\chi>1$ is the pure (anti)kaon condensed phase. In the mixed phase, the total energy density also changes the form to,
\begin{equation} \label{eqn.27}
    \varepsilon = (1-\chi) \varepsilon^h + \chi \varepsilon^{\bar{K}}
\end{equation}
where, $\varepsilon^h$, $\varepsilon^{\bar{K}}$ denotes the total energy density in hadronic and kaon phases respectively.

\subsection{Coupling parameters}

In the NL CDF model, we adopt the GMT \cite{2000NuPhA.674..553P} and NL3 \cite{1997PhRvC..55..540L} parametrizations for meson-nucleon couplings. For the DDRH model, DD-ME2  \cite{2005PhRvC..71b4312L}, DD2 \cite{2010PhRvC..81a5803T} and PKDD \cite{2004PhRvC..69c4319L} parametrizations are considered for meson-nucleon couplings. Table-\ref{tab:1} provides the NL coupling parameters implemented in this work. 
\begin{table} [h!]
\centering
\caption{The coupling constants for the CDF models GMT, NL3, at nuclear saturation density $n_0$.}
\begin{tabular}{cccccc}
\hline \hline
Model & $g_{\sigma N}$ & $g_{\omega N}$ & $g_{\rho N}$ & $g_2$ (fm$^{-1}$) & $g_3$ \\
\hline
GMT & 9.9400 & 12.2981 & 9.2756 & 10.5745 & $-24.1907$ \\
NL3 & 10.2170 & 12.8680 & 8.9480 & 10.4310 & $-28.8850$ \\
\hline
\end{tabular}
\label{tab:1}
\end{table}

In case of the density-dependent model, the meson-nucleon coupling constants are functions of density as
\begin{equation}\label{eqn.30}
g_{i N}(n)= g_{i N}(n_{0}) f_i(x) \quad \quad \text{for }i=\sigma,\omega
\end{equation}
where, $x=n/n_0$, $n_0$ is the nuclear saturation density and
\begin{equation}\label{eqn.31}
f_i(x)= a_i \frac{1+b_i (x+d_i)^2}{1+c_i (x+d_i)^2}
\end{equation}
For the $\rho$-meson, the density-dependent coupling constant is given by
\begin{equation}\label{eqn.32}
g_{\rho N}(n)= g_{\rho N}(n_{0}) e^{-a_{\rho}(x-1)}
\end{equation}
The density-dependent hadronic model parameters implemented in this work are listed in table-\ref{tab:11}.

\begin{table} [h!]
\centering
\caption{The coupling constants for the DDRH models DD-ME2, DD2, PKDD, at $n_0$.}
\begin{tabular}{cccc}
\hline \hline
Model & $g_{\sigma N}$ & $g_{\omega N}$ & $g_{\rho N}$ \\
\hline
DD-ME2 & 10.5396 & 13.0189 & 7.3672 \\
DD2 & 10.686681 & 13.342362 & 7.25388 \\
PKDD & 10.7385 & 13.1476 & 8.5996 \\
\hline
\end{tabular}
\label{tab:11}
\end{table}

\begin{table*} [t!]
\centering
\caption{The coefficient values for the density-dependent coupling parametrizations, at $n_0$.}
\begin{tabular}{cccccccccc}
\hline \hline
Model & $a_{\sigma}$ & $a_{\omega}$ & $a_{\rho}$ & $b_{\sigma}$ & $b_{\omega}$ & $c_{\sigma}$ & $c_{\omega}$ & $d_{\sigma}$ & $d_{\omega}$ \\
\hline
DD-ME2 & 1.3881 & 1.3892 & 0.5647 & 1.0943 & 0.9240 & 1.7057 & 1.4620 & 0.4421 & 0.4775 \\
DD2 & 1.35763 & 1.369718 & 0.518903 & 0.634442 & 0.496475 & 1.005358 & 0.817753 & 0.57581 & 0.638452 \\
PKDD & 1.327423 & 1.342170 & 0.183305 & 0.435126 & 0.371167 & 0.691666 & 0.611397 & 0.694210 & 0.738376 \\
\hline
\end{tabular}
\label{tab:5}
\end{table*}
The bare nucleon masses are considered to be identical for both the NL mean field model cases as $m_N = 939$ MeV. In the DDRH models we consider  $m_N=938.9,~938.565,~939.573$ MeV for DD-ME2, DD2, PKDD parametrizations respectively. The bare mass of the (anti)kaons in this calculation is considered to be 493.69 MeV.

The coefficients in eqs.\eqref{eqn.31}-\eqref{eqn.32} are fixed by several procedures. For details we refer the readers to the references.-\cite{2010PhRvC..81a5803T,2005PhRvC..71b4312L, 2004PhRvC..69c4319L}. The parameter values employed in the DDRH model to get the density dependence is given in table-\ref{tab:5}. The meson-masses for the GMT, NL3, DD-ME2, DD2, PKDD models  of the coupling constants incorporated in the calculations  are listed in table-\ref{tab:4}.
\begin{table} [h!]
\centering
\caption{Mass values (in units of MeV) of the $\sigma$, $\omega$, $\rho$-mesons in GMT, NL3, DD-ME2, DD2, PKDD parametrizations.}
\begin{tabular}{cccc}
\hline \hline
Model & $m_\sigma$ & $m_\omega$ & $m_\rho$ \\
\hline
GMT & 511.198 & 783.000 & 770.000 \\
NL3 & 508.194 & 782.501 & 763.000 \\
DD-ME2 & 550.124 & 783.000 & 763.000 \\
DD2 & 546.212459 & 783.000 & 763.000 \\
PKDD & 555.5112 & 783.000 & 763.000 \\
\hline
\end{tabular}
\label{tab:4}
\end{table}

The respective parametrizations satisfy the values of the quantities at $n_0$ as shown in the table-\ref{tab:10}. In the table, $E/A$, $K_0$, $a_{sym}$ denote the binding energy per nucleon, compression modulus, symmetry energy coefficient.
\begin{table} [h!]
\centering
\caption{The nuclear properties of the CDF models at respective nuclear saturation densities.}
\begin{tabular}{cccccc}
\hline \hline
Model & $n_0$ & $E/A$ & $K_0$ & $a_{sym}$ & $m^*_N/m_N$ \\
 & (fm$^{-3}$) & (MeV) & (MeV) & (MeV) & \\
\hline
GMT & 0.145 & $-16.3$ & 281.00 & 36.90 & $0.634$ \\
NL3 & 0.148 & $-16.299$ & 271.76 & 37.40 & $0.600$ \\
DD-ME2 & 0.152 & $-16.14$ & 250.89 & 32.30 & 0.572 \\
DD2 & 0.149065 & $-16.02$ & 242.70 & 32.73 & 0.5625 \\
PKDD & 0.149552 & $-16.267$ & 262.181 &  36.79 &  0.5712 \\
\hline
\end{tabular}
\label{tab:10}
\end{table}

The meson-(anti)kaon couplings are not considered to be density-dependent \cite{2014PhRvC..90a5801C} in DDRH model. The vector coupling parameters in the kaon sector are evaluated from the iso-spin counting rule and quark model \cite{2001PhRvC..64e5805B} as
\begin{equation}\label{eqn.28}
    g_{\omega K} = \frac{1}{3} g_{\omega N}, \quad  g_{\rho K} = g_{\rho N}
\end{equation}
and for the scalar coupling parameters, they are calculated at nuclear saturation density from the real part of $K^-$ optical potential depth as
\begin{equation} \label{eqn.29}
    U_{\bar{K}} (n_0) = - g_{\sigma K} \sigma (n_0) - g_{\omega K} \omega_0 (n_0) + \Sigma^r_{N} (n_0)
\end{equation}
where, $\Sigma^r_{N} (n_0)$ is the contribution from the nucleons alone and is not considered in case of NL CDF model.

Experimental studies \cite{1997NuPhA.625..372L,2000PhRvC..62f1903P} show that the kaons experience a repulsive interaction in nuclear matter whereas antikaons experience an attractive potential. Several model calculations \cite{1994PhLB..337....7K,1997NuPhA.625..287W,1998PhLB..426...12L, 2000NuPhA.671..481R,2000NuPhA.669..153S} provide a very broad range of optical potential values as $-120 \leq U_{\bar{K}} \leq -40$ MeV. Another calculation from hybrid model \cite{1999PhRvC..60b4314F} suggests the value of $K^-$ optical potential to be in the range $180 \pm 20$ MeV at nuclear saturation density. In this work, we have considered a $K^-$ potential range of $-160 \leq U_{\bar{K}} \leq -120$ MeV and the scalar meson-(anti)kaon couplings evaluated for the above potential depth range are listed in table-\ref{tab:2}.
\begin{table} [h!]
\centering
\caption{Parameter values of the scalar $\sigma$ meson-(anti)kaon couplings in GMT, NL3, DD-ME2 and DD2 parametrizations at $n_0$.}
\begin{tabular}{cccc}
\hline \hline
Model & \multicolumn{3}{c}{$U_{\bar{K}}$ (MeV)} \\
 & $-120$ & $-140$ & $-160$ \\
\hline
GMT & 0.8217 & 1.4006 & 1.9796 \\
NL3 & 0.4707 & 1.0088 & 1.5469 \\
DD-ME2 & 0.4311 & 0.9553 & 1.4796 \\
DD2 & 0.3155 & 0.8359 & 1.3562 \\
PKDD & 0.4309 & 0.9639 & 1.4970 \\
\hline
\end{tabular}
\label{tab:2}
\end{table}

\section{Results} \label{sec:results}

We model the matter to be composed of nucleons with the possibility of (anti)kaon condensate appearance with increasing density. We compare the matter properties and the resultant star structure with different approach in calculating the EOS of the matter within relativistic mean field theory considering different (anti)kaon optical potentials $U_{\bar{K}}= -120, -140, -160$ MeV. 

\subsection{Non-Linear CDF model}

\begin{figure*} [t!]
  \begin{center}
\includegraphics[width=14cm,keepaspectratio ]{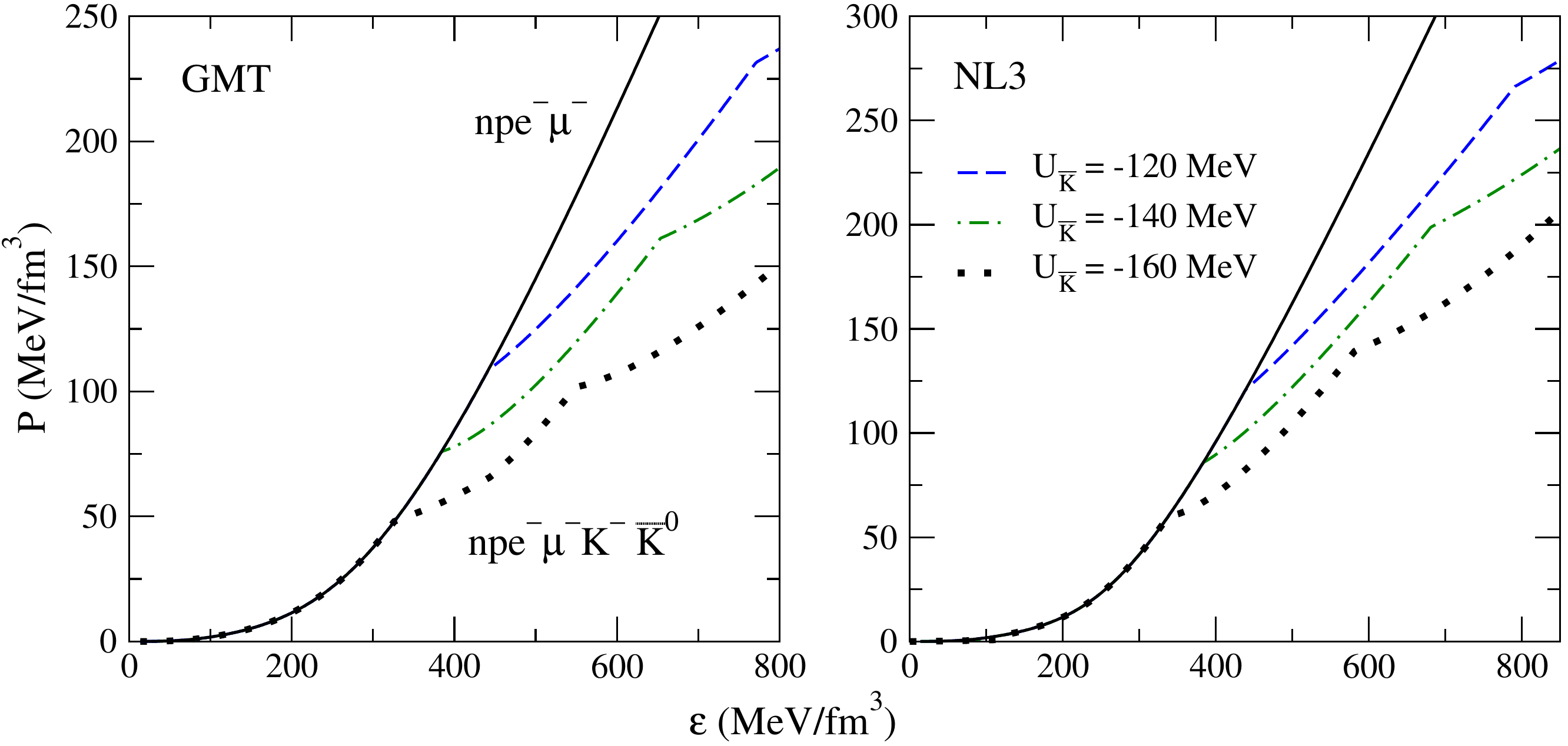}
\caption{The variation of matter pressure with energy density. Left panel: GMT parameterization, right panel: NL3 parameterization cases with several (anti)kaon potential depths ($U_{\bar{K}}$). The solid lines represent the composition of only nucleons ($n,p$) and leptons ($e^-,\mu^-$). The other curves represent the composition with (anti)kaons in addition to nucleons and leptons. The dashed curve depicts $U_{\bar{K}}=-120$ MeV, dash-dotted curve with $U_{\bar{K}}=-140$ MeV and dotted curve with $U_{\bar{K}}=-160$ MeV.}
\label{fig-4}
\end{center}
\end{figure*}

In the case of GMT parametrization, the transition from hadronic to kaonic phase is observed to be of second order for $U_{\bar{K}}=$ $-120,-140$ MeV but of first order for $U_{\bar{K}}=-160$ MeV. While in case of NL3 parameterization, the phase transition is of second order for the whole range of  $K^-$ potential adopted here. 

Fig.\ref{fig-4} shows the matter pressure as a function of energy density for both the GMT and NL3 models. The appearance of (anti)kaons to a great extent softens the EOS. The two kinks in the EOSs marks the onset of $K^-$ and $\bar{K}^0$ respectively. The two kinks are observed to be in higher densities for the NL3 model in comparison to the GMT case referring the delay of (anti)kaons into the matter for the former case. The appearance of (anti)kaon condensation is through first order transition only for $K^-$ with $U_{\bar{K}}=-160$ MeV in GMT parametrization. In other cases it is second order phase transition.

\begin{figure} [h!]
  \begin{center}
\includegraphics[width=8.6cm,keepaspectratio ]{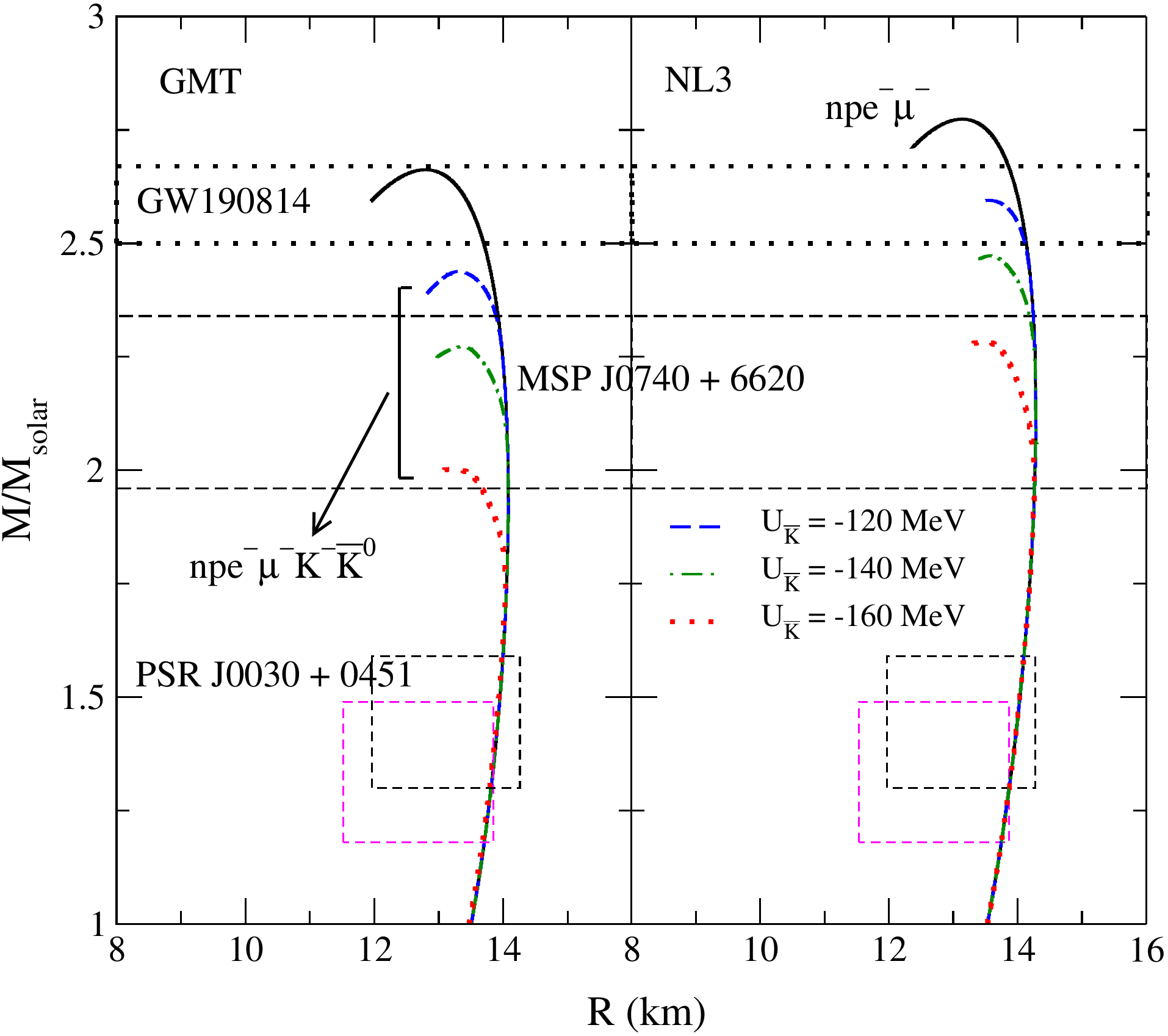}
\caption{The M-R relations corresponding to the EOSs shown in fig.\ref{fig-4}. Left panel: GMT, right panel: NL3 parameterization. The solid curve corresponds to the pure nucleonic star, while the dashed, dash-dotted, dotted curves correspond to stars with (anti)kaon condensation and optical potentials, $U_{\bar{K}}=-120, -140, -160$ MeV respectively. The mass constraints from the various astrophysical observations are represented by the region enclosed by dotted corresponding to the GW190814 observation~\cite{2020ApJ...896L..44A}, dashed-lines to the  MSP J0740+6620 \cite{2020NatAs...4...72C}. The mass-radius limits obtained for PSR J0030+0451 from the NICER experiment \cite{2019ApJ...887L..24M,2019ApJ...887L..21R} are represented by the squared dashed regions.}
\label{fig-12}
\end{center}
\end{figure}

Fig.\ref{fig-12} shows the results of the mass-radius (M-R) relationship for static spherical stars from solution of the Tolman-Oppenheimer-Volkoff (TOV) equations \cite{1996cost.book.....G} corresponding to the EOSs discussed here and shown in fig.\ref{fig-4}. For the crust, we have considered the EOS of Baym, Pethick and Sutherland \cite{1971ApJ...170..299B}. Table-\ref{tab:9} provides the set of maximum mass values, corresponding radius and central density for the nucleons and (anti)kaon EOSs with various values of $U_{\bar{K}}$. Inclusion of (anti)kaons leads to reduction of maximum mass of the compact stars.

\begin{table} [h!]
\centering
\caption{Parameter values of the maximum mass stars. Here, maximum mass, M$_{max}$ (in units of M$_\odot$), radius (in km), corresponding central density $n_c$ (in units of $n_0$) of nucleon compact stars for different values of $K^-$ optical potential depths $U_{\bar{K}}$ (in units of MeV) at $n_0$ in GMT and NL3 parametrization models.}
\begin{tabular}{c|ccc|ccc}
\hline \hline
 & \multicolumn{3}{c}{GMT} & \multicolumn{3}{|c}{NL3} \\
 \cline{2-7}
$U_{\bar{K}}$ & M$_{max}$ & R & $n_c$ & M$_{max}$ & R & $n_c$ \\
(MeV) & (M$_{\odot}$) & (km) & ($n_0$) & (M$_{\odot}$) & (km) & ($n_0$) \\
\hline
$0$ & 2.66 & 12.80 & 4.91 & 2.77 & 13.14 & 4.52 \\
$-120$ & 2.44 & 13.31 & 4.76 & 2.59 & 13.56 & 4.37 \\
$-140$ & 2.27 & 13.36 & 4.75 & 2.47 & 13.59 & 4.47 \\
$-160$ & 2.00 & 13.22 & 4.78 & 2.28 & 13.48 & 4.53 \\
\hline
\end{tabular}
\label{tab:9}
\end{table}

\begin{figure} [t!]
\includegraphics[width=8.5cm,keepaspectratio ]{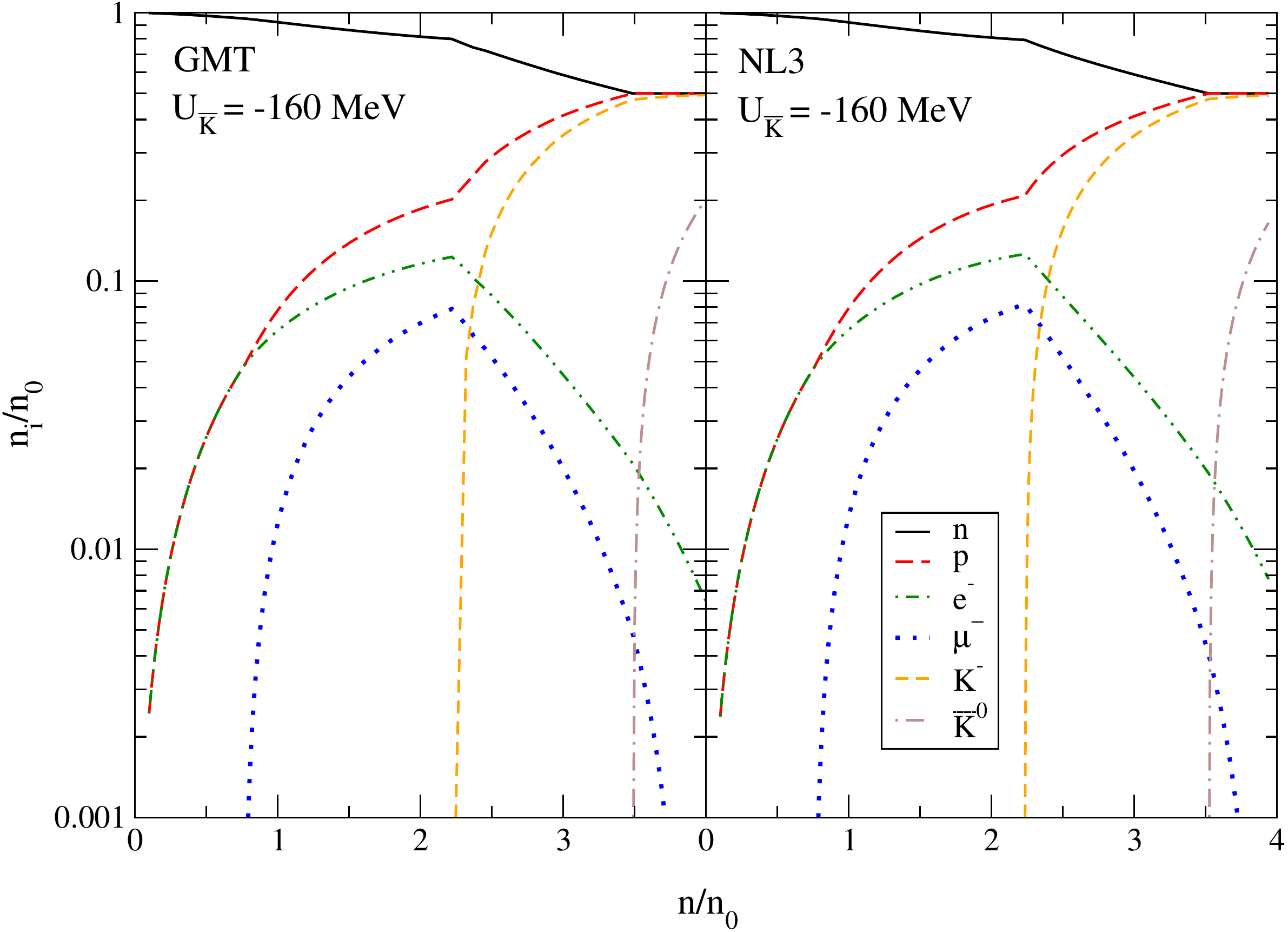}
\caption{Population densities $n_i$ (in units of $n_0$) of various species as a function of baryon number density. Left panel: GMT, right panel: NL3 model and $K^-$ potential depth of $-160$ MeV. Solid lines denote neutron ($n$), long-dashed lines proton ($p$), dash-dotted curves electron ($e^-$), dotted lines muon ($\mu^-$), short-dashed lines $K^-$ and dash-double dotted lines denote $\bar{K}^0$ population.}
\label{fig-1}
\end{figure}

The variation of particle fractions in the matter with total baryonic number density is shown in the fig.\ref{fig-1} for $U_{\bar{K}} (n_0)=-160$ MeV for both the parametrization sets. For GMT parametrization the mixed phase initiates with the onset of $K^-$ at $\sim 2.22~n_0$. In addition to the global conservation rule of baryon number (eq.\eqref{eqn.25}) and charge neutrality (eq.\eqref{eqn.26}), pressure and chemical potential equilibrium conditions between two phases (eqs.\eqref{eqn.23},\eqref{eqn.24}) determines the mixed phase region. Due to higher rest mass of (anti)kaons compared to the lepton species and being bosons, the former condense in the lowest energy state and so are preferred to maintain the global charge neutrality condition. This results in decrease in electron and muon populations as can be clearly visualized in fig.\ref{fig-1}. The mixed phase terminates at $\sim 2.90~n_0$. Further, with the appearance of $\bar{K}^0$ at $\sim 3.49~n_0$ and ceasing of electron population around $4-4.5~n_0$ the proton and $K^-$ populations becomes equal following the charge neutrality condition. $\bar{K}^0$ condensates through the second order phase transition. However, for NL3 parametrization the phase transition occurs via second order for both the (anti)kaons ($K^-,\bar{K}^0$) implying the absence of mixed phase regime. Even though we are fixing the $U_{\bar{K}}$ identical to GMT model cases, the (anti)kaons appear at a slightly higher densities compared to the former case.

The threshold densities for the onset of $K^-,\bar{K}^0$ in the dense nuclear matter for different $K^-$ potentials are provided in Table-\ref{tab:3}. It can be observed that the threshold densities shifts towards lower densities with the increase in strength of $U_{\bar{K}}$ at $n_0$.

\begin{table} [h!]
\centering
\caption{Threshold densities, $n_{cr}$ (in units of $n_0$) for antikaon condensation in dense nuclear matter for different values of $K^-$ optical potential depths $U_{\bar{K}}$ (in units of MeV) at $n_0$.}
\begin{tabular}{c|cc|cc}
\hline \hline
 & \multicolumn{2}{c}{GMT} & \multicolumn{2}{|c}{NL3} \\
 \cline{2-5}
$U_{\bar{K}}$ & $n_{cr}$($K^-$) & $n_{cr}$($\bar{K}^0$) & $n_{cr}$($K^-$) & $n_{cr}$($\bar{K}^0$) \\
(MeV) & ($n_0$) & ($n_0$) & ($n_0$) & ($n_0$) \\
\hline
$-120$ & 2.87 & 4.45 & 2.77 & 4.35 \\
$-140$ & 2.56 & 3.96 & 2.49 & 3.94 \\
$-160$ & 2.22 & 3.49 & 2.24 & 3.53 \\
\hline
\end{tabular}
\label{tab:3}
\end{table}

\begin{figure} [h!]
\includegraphics[width=8.5cm,keepaspectratio ]{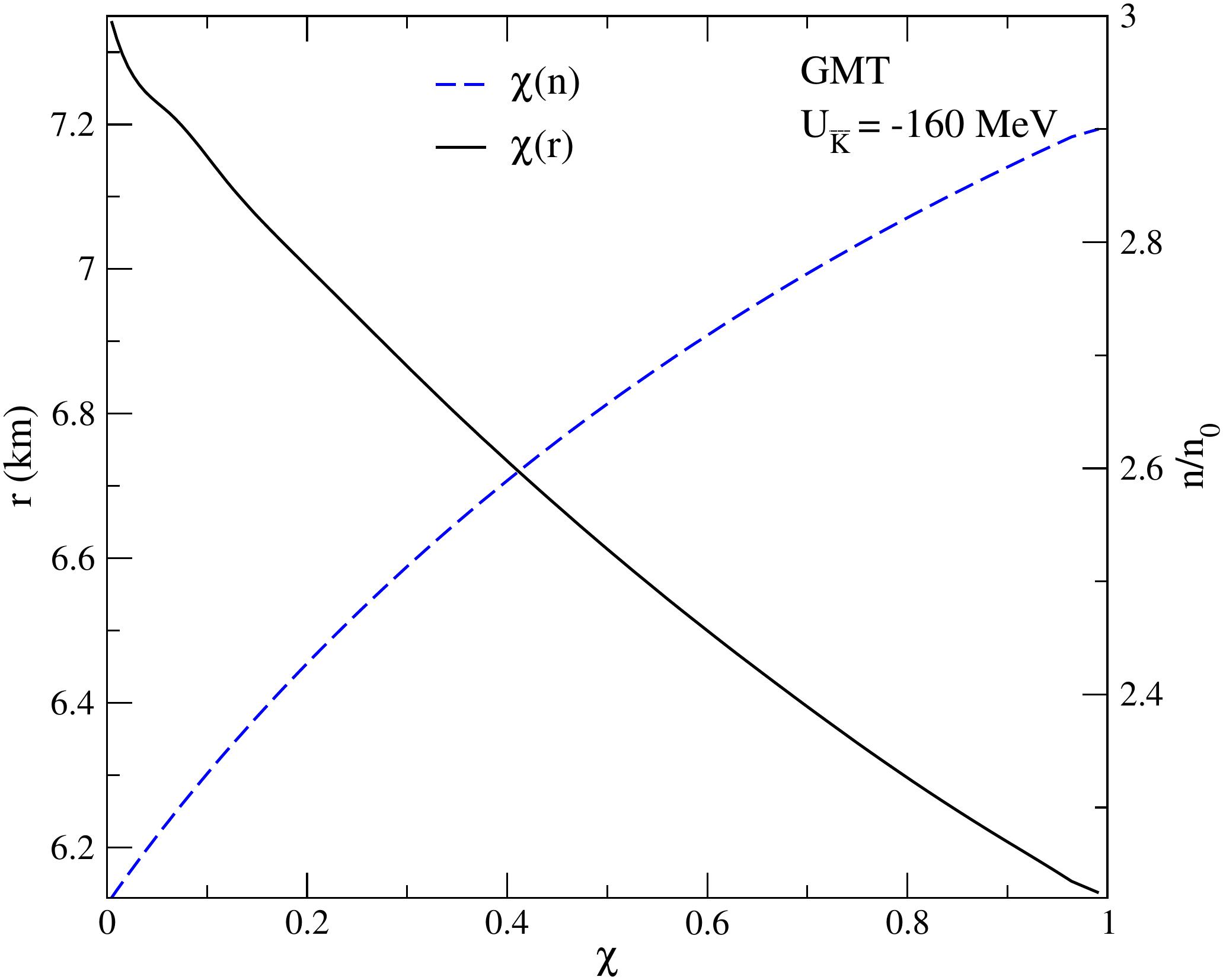}
\caption{Baryon number density ($n$), radial distance ($r$) inside the compact star as a function of volume fraction of the (anti)kaon condensate ($\chi$) for the GMT model case with $U_{\bar{K}}=-160$ MeV. The solid curve denotes $\chi(r)$, while dashed curve represents $\chi(n)$.}
\label{fig-13}
\end{figure}

Fig.\ref{fig-13} shows the extent of mixed phase region inside the neutron star modelled with GMT parametrization. It is evident that the mixed phase regime starts ($\chi\sim 0$) from around matter density of $\sim 2.22~n_0$ which corresponds to star radius $7.34$ km, and terminates ($\chi \sim 1$) at around matter density of $2.9~n_0$ or corresponding stellar radius of, $\sim 6.13$ km.

\begin{figure} [h!]
\includegraphics[width=8.3cm,keepaspectratio ]{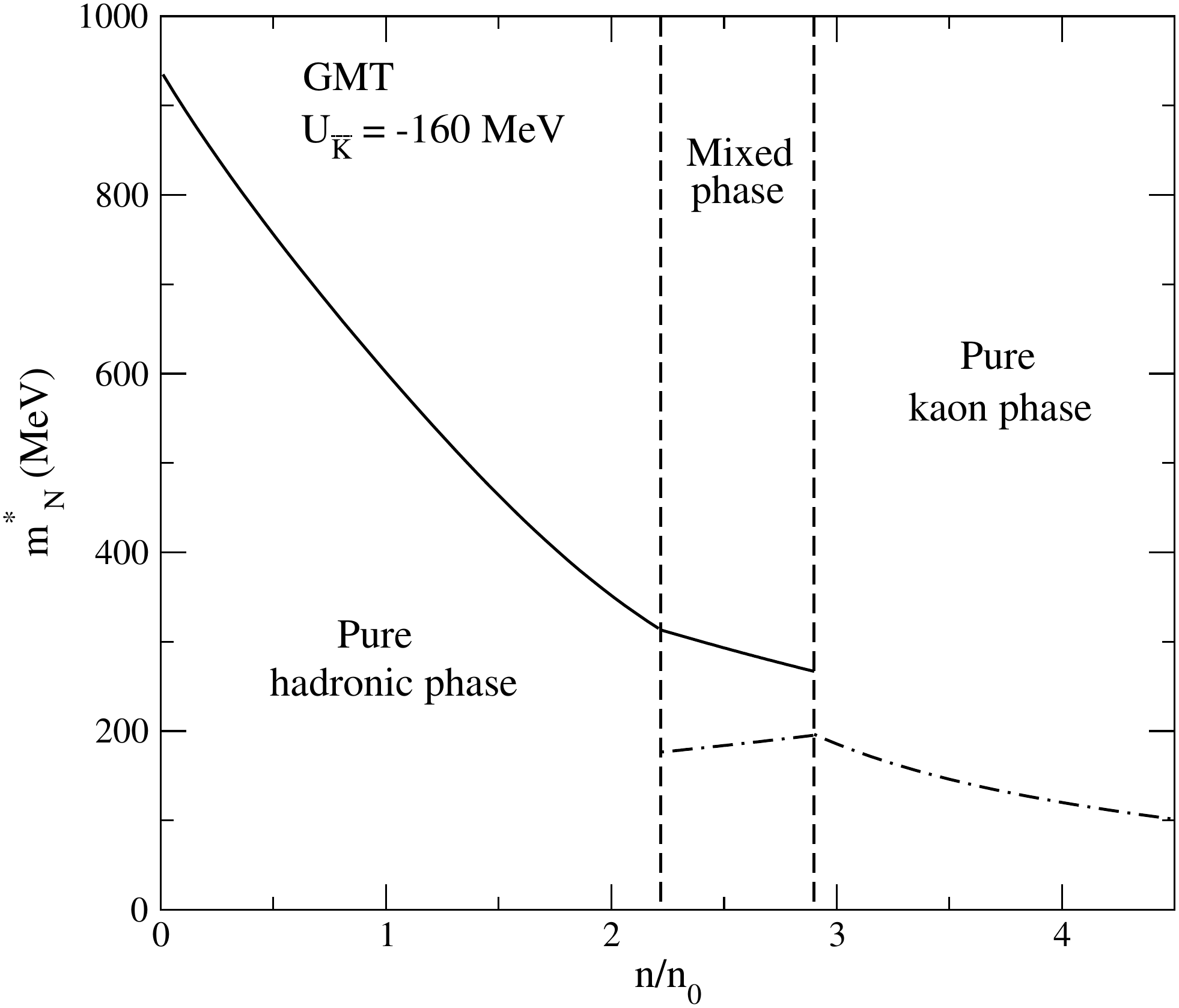}
\caption{Effective nucleon mass, $m^{*}_N$ as a function of baryon number density for the GMT model case with $U_{\bar{K}}=-160$ MeV. The solid curve denotes $m^{*}_N (n_B)$ in hadronic phase, while dash-dotted curve represents $m^{*}_N (n_B)$ in kaonic phase. Region with $n<2.2~n_0$ is the pure hadronic phase, between $2.2\leq n \leq 2.9~n_0$ is the mixed phase and $n>2.9~n_0$ is the pure (anti)kaonic phase.}
\label{fig-18}
\end{figure}

The interactions between (anti)kaons and nucleons alters the nucleon effective mass in the mixed phase regime where both the hadronic and kaonic phase co-exist. This effect in shown in fig.\ref{fig-18}. With a difference of $\sim 100$ MeV, the effective nucleon masses are observed to increase in kaonic phase while it decreases in the pure hadronic phase as we move interior towards the pure kaonic phase regime.

\begin{figure} [h!]
\includegraphics[width=8.5cm,keepaspectratio ]{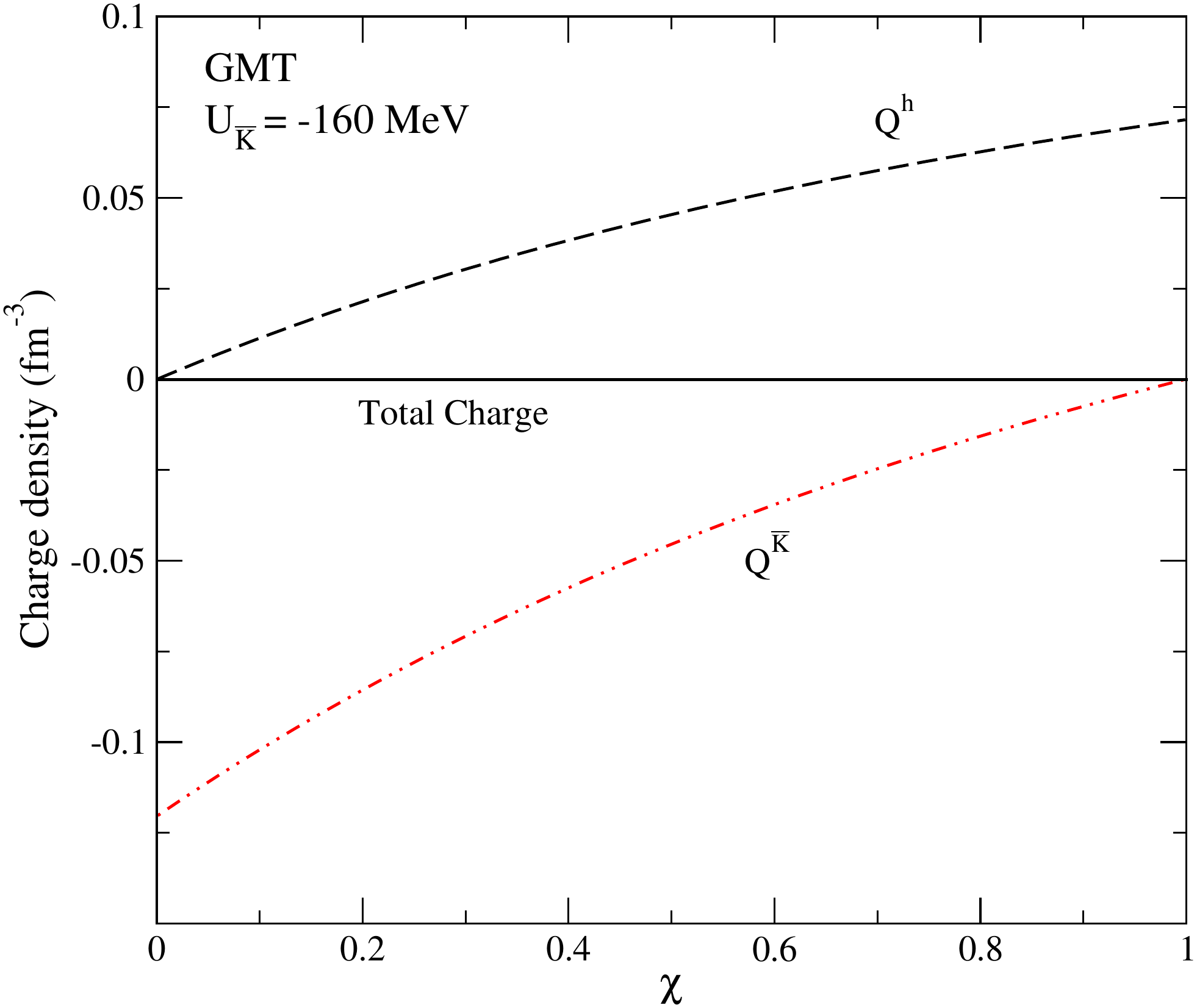}
\caption{Charge densities in the pure hadronic phase and (anti)kaon condensed phase as a function of volume fraction of the latter for the GMT model case with $U_{\bar{K}}=-160$ MeV. $Q^h, Q^{\bar{K}}$ denote the charge in hadronic and kaonic phases respectively.}
\label{fig-3}
\end{figure}

The charge densities of each normal and kaon condensed phase in the mixed phase region for the GMT model with $U_{\bar{K}}=-160$ MeV as a function of the kaon volume fraction is shown in fig.\ref{fig-3}. The central solid black line represents the global charge neutrality condition as provided in eq.\eqref{eqn.26}.

\begin{figure} [h!]
  \begin{center}
\includegraphics[width=8.5cm,keepaspectratio ]{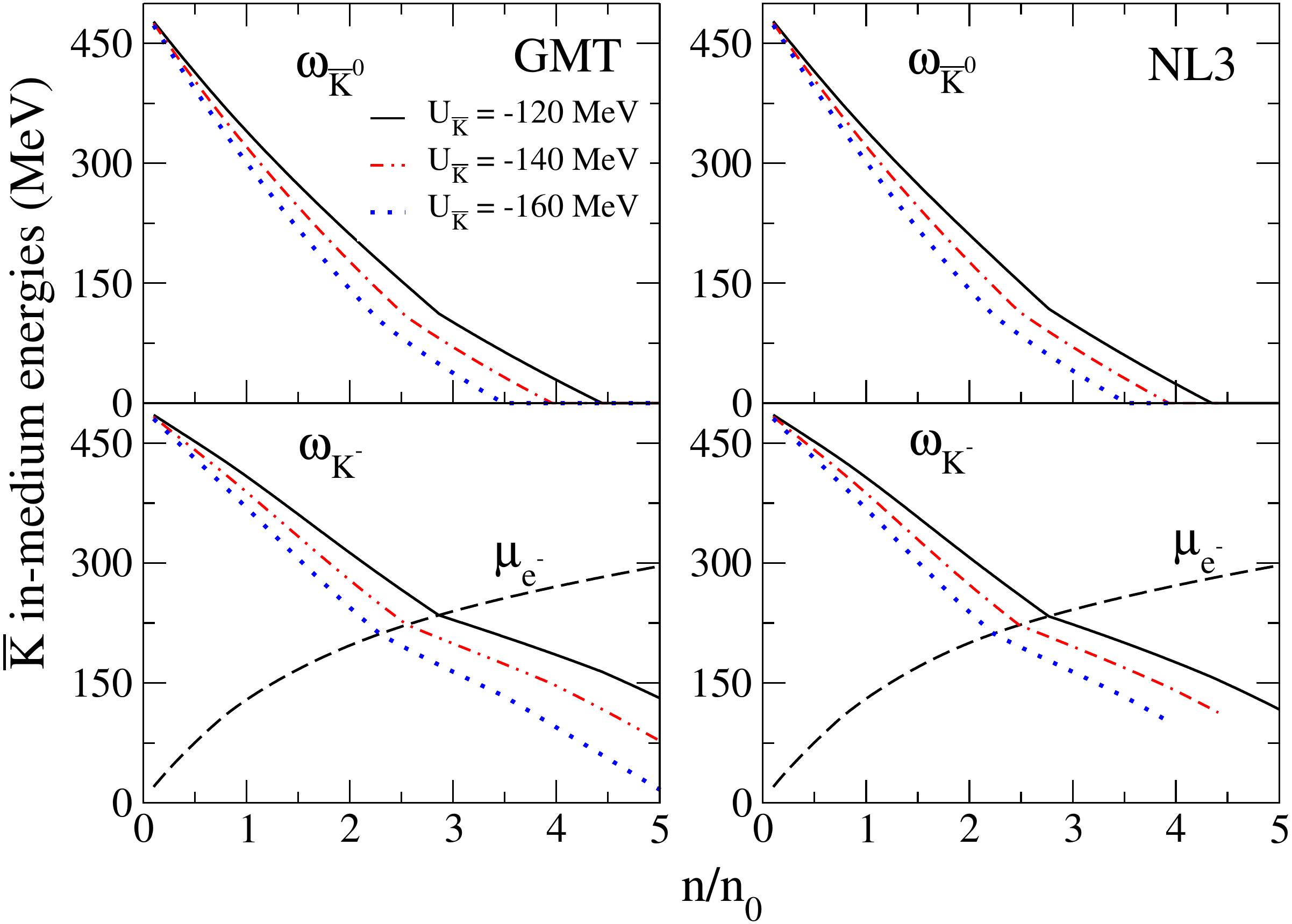}
\caption{The effective energy of (anti)kaons as a function of baryon number density, left panels: GMT, right panels: NL3 parametrization. Upper panels: in-medium energies of $\bar{K}^0$, lower panels: in-medium energies of $K^-$ with several $U_{\bar{K}}$. The dashed lines represent the respective electron chemical potentials for each model case. The solid curve exhibits the $U_{\bar{K}}$ strengths of $-120$ MeV, dash-dotted lines exhibits $-140$ MeV case and dotted lines exhibits the $-160$ MeV case.}
\label{fig-5}
\end{center}
\end{figure}

The (anti)kaon energies as a function of baryon number density with $U_{\bar{K}}=-120$, $-140$, $-160$ MeV for the two parameter sets (GMT, NL3) are shown in fig.\ref{fig-5}. The in-medium energies for both the (anti)kaons decrease with density. The dashed curve representing $\mu_e$ crossing over the $\omega_{\bar{K}}$ curves marks the end of pure hadronic phase and initiation of pure kaonic phase. $K^-$ condensations initiates once the value of $\omega_{K^-}$ reaches that of the electron chemical potential and $\bar{K}^0$ condensation starts when the value of $\omega_{\bar{K}^0}$ equates to zero. From fig.\ref{fig-5}, it is observed that the threshold condition, $\omega_{K^-}=\mu_e$ is achieved way early than the $\omega_{\bar{K}^0}=0$ one, leading to earlier appearance of $K^-$.

\begin{figure} [h!]
\includegraphics[width=8.5cm,keepaspectratio ]{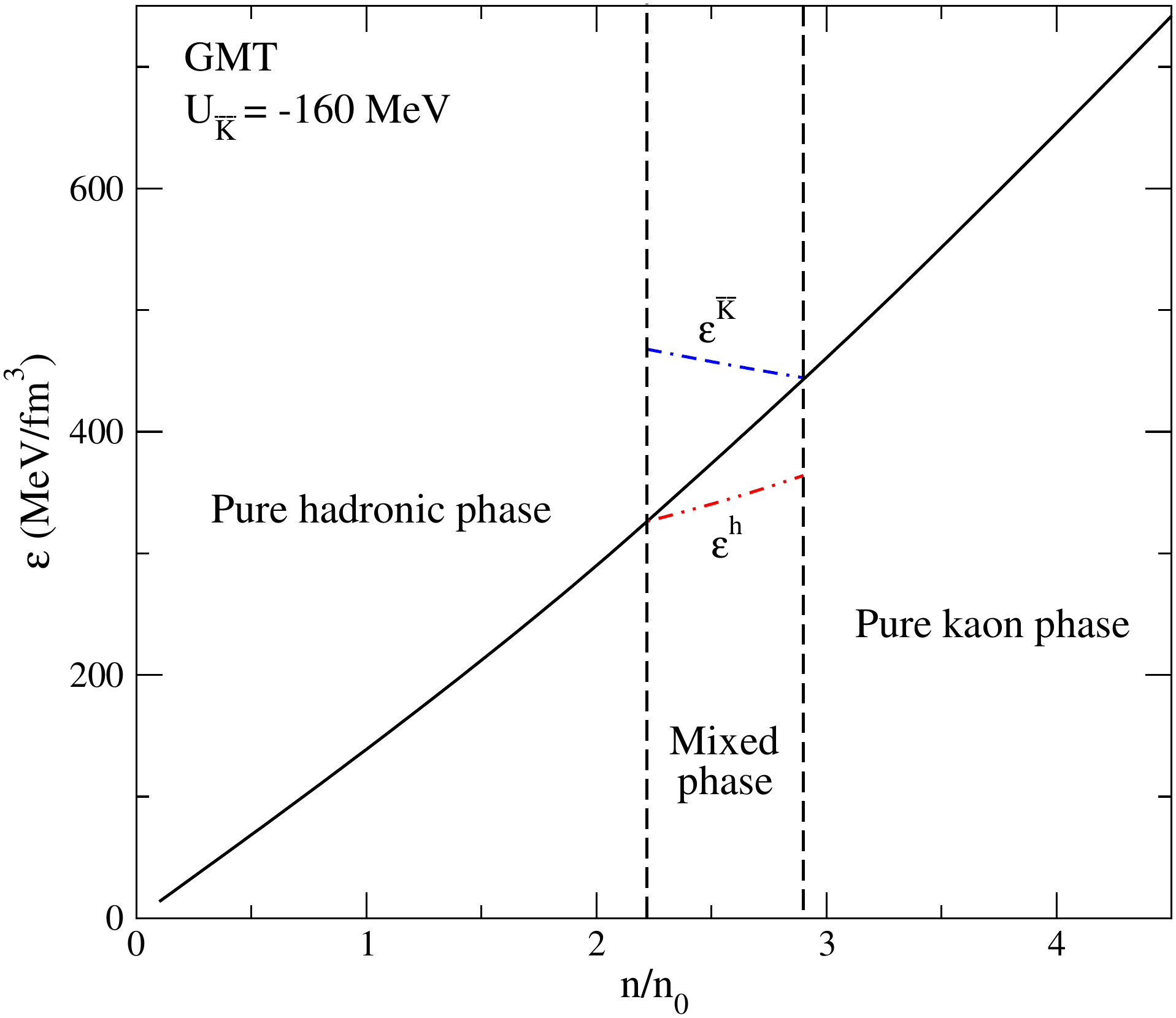}
\caption{Energy density as a function of baryon number density for the GMT model with $U_{\bar{K}}=-160$ MeV for the $3$ phases. The solid curve denotes the total energy density variation, while dash-dotted curves represent $\varepsilon^h$ and dash-double dotted curve represent $\varepsilon^{\bar{K}}$. The three regions demarcation are similar as fig.\ref{fig-18}.}
\label{fig-14}
\end{figure}

The difference in the energy densities of the hadronic and kaonic phases for GMT parameterization with $U_{\bar{K}}=-160$ MeV is shown in fig.\ref{fig-14}. The total energy density in the mixed phase region is evaluated from eq.\eqref{eqn.27} and grows monotonically with density.

\begin{figure*} [t!]
  \begin{center}
\includegraphics[width=15.5cm,keepaspectratio ]{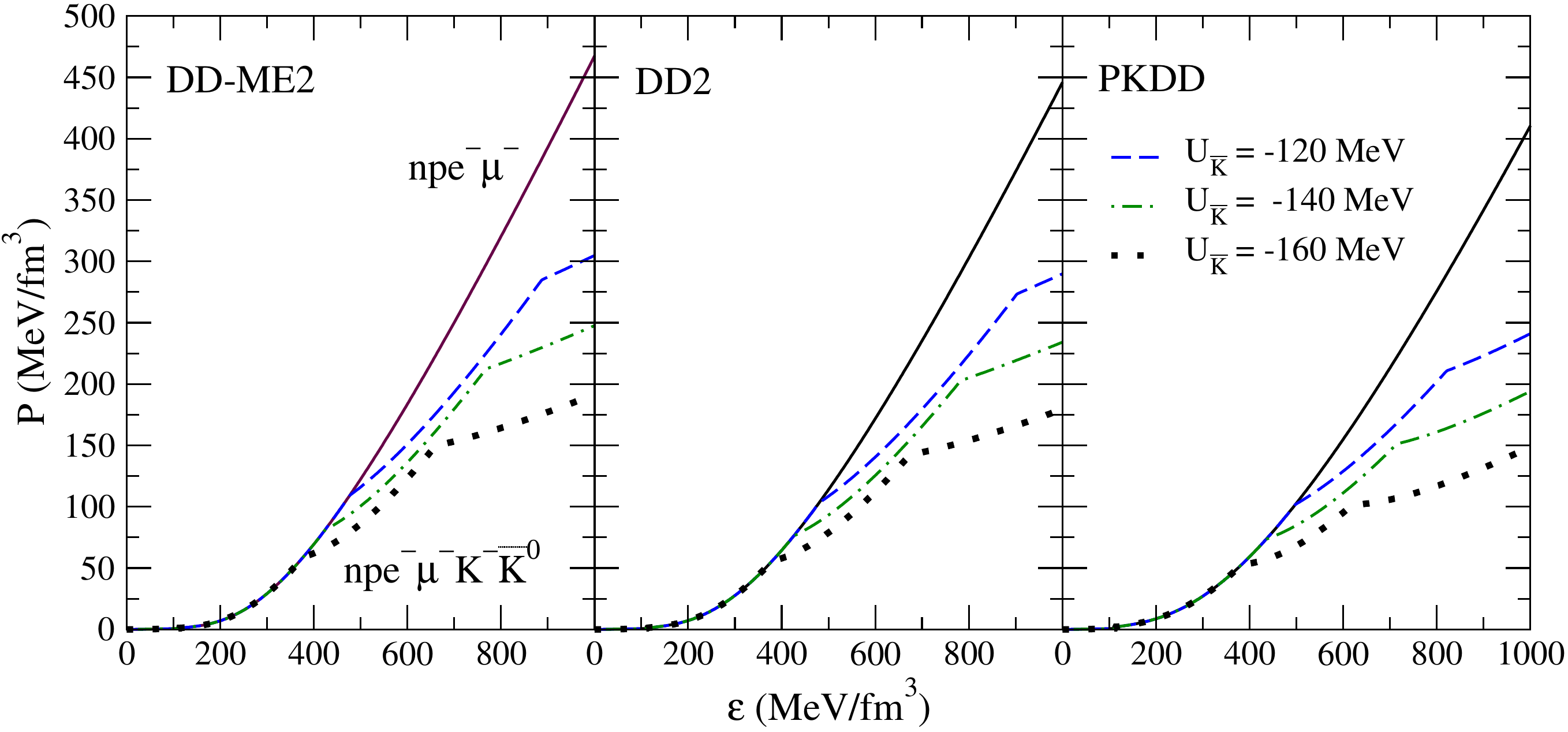}
\caption{The equation of states, Left panel: DD-ME2, center panel: DD2, right panel: PKDD parametrization case with several (anti)kaon potential depths ($U_{\bar{K}}$). The solid lines represent the composition of only nucleons ($n,p$) and leptons ($e^-,\mu^-$). The dashed curves depict $U_{\bar{K}}=-120$ MeV, dash-dotted curves with $U_{\bar{K}}=-140$ MeV and dotted curves with $U_{\bar{K}}=-160$ MeV represent the composition with (anti)kaons.}
\label{fig-15}
\end{center}
\end{figure*}

\begin{table*} [t!]
\centering
\caption{Threshold densities, $n_{cr}$ (in units of $n_0$) for antikaon condensation in dense nuclear matter for different values of $K^-$ optical potential depths $U_{\bar{K}}$ (in units of MeV) at $n_0$ with density-dependent DD-ME2, DD2, PKDD parametrizations.}
\begin{tabular}{c|cc|cc|cc}
\hline \hline
 & \multicolumn{2}{c}{DD-ME2} & \multicolumn{2}{|c}{DD2} & \multicolumn{2}{|c}{PKDD} \\
 \cline{2-7}
 $U_{\bar{K}}$ (MeV) & $n^{K^-}_{cr}$($n_0$) & $n^{\bar{K}^0}_{cr}$($n_0$) & $n^{K^-}_{cr}$($n_0$) & $n^{\bar{K}^0}_{cr}$($n_0$) & $n^{K^-}_{cr}$($n_0$) & $n^{\bar{K}^0}_{cr}$($n_0$) \\
\hline
 $-120$ & 3.00 & 4.89 & 3.08 & 5.11 & 3.16 & 4.75 \\
 $-140$ & 2.72 & 4.42 & 2.79 & 4.62 & 2.82 & 4.29 \\
 $-160$ & 2.47 & 3.96 & 2.53 & 4.14 & 2.53 & 3.84 \\
\hline
\end{tabular}
\label{tab:7}
\end{table*}

\subsection{Density-Dependent CDF model}

For DDRH model, the phase transition from hadronic to kaonic phase for all parametrizations and all (anti)kaon optical potential depths considered in the present work is through the second-order phase transition.

The matter pressure as a function of energy density (EOS) for DDRH model is shown in fig.\ref{fig-15}. It is observed that the stiffest EOS results in for the DD-ME2 parametrization. Incorporation of $\bar{K}$ softens the EOS and more pronounced effects are seen with deeper $K^-$ optical potential depths. The two kinks in the EOS marks the onsets of $K^-$ and $\bar{K}^0$ respectively. While the first kink appears to be $\sim 400$ MeV/fm$^3$ for all the three parametrizations, the onset of $\bar{K}^0$ denoted by the second kink is delayed the most in DD2 parametrization case.

Table-\ref{tab:7} provides the threshold densities of (anti)kaon condensation in different potential depths of (anti)kaons. It can be observed from table-\ref{tab:7} that the appearance of $K^{-}$-mesons is foremost in case of DD-ME2 model irrespective of the optical potential. On the other hand, the appearance of $\bar{K}^0$ is the earliest in PKDD model among others. With the increasing potential depth of (anti)kaons in symmetric nuclear matter, the onset of (anti)kaons happens earlier.

\begin{figure*} [t!]
  \begin{center}
\includegraphics[width=15.5cm,keepaspectratio ]{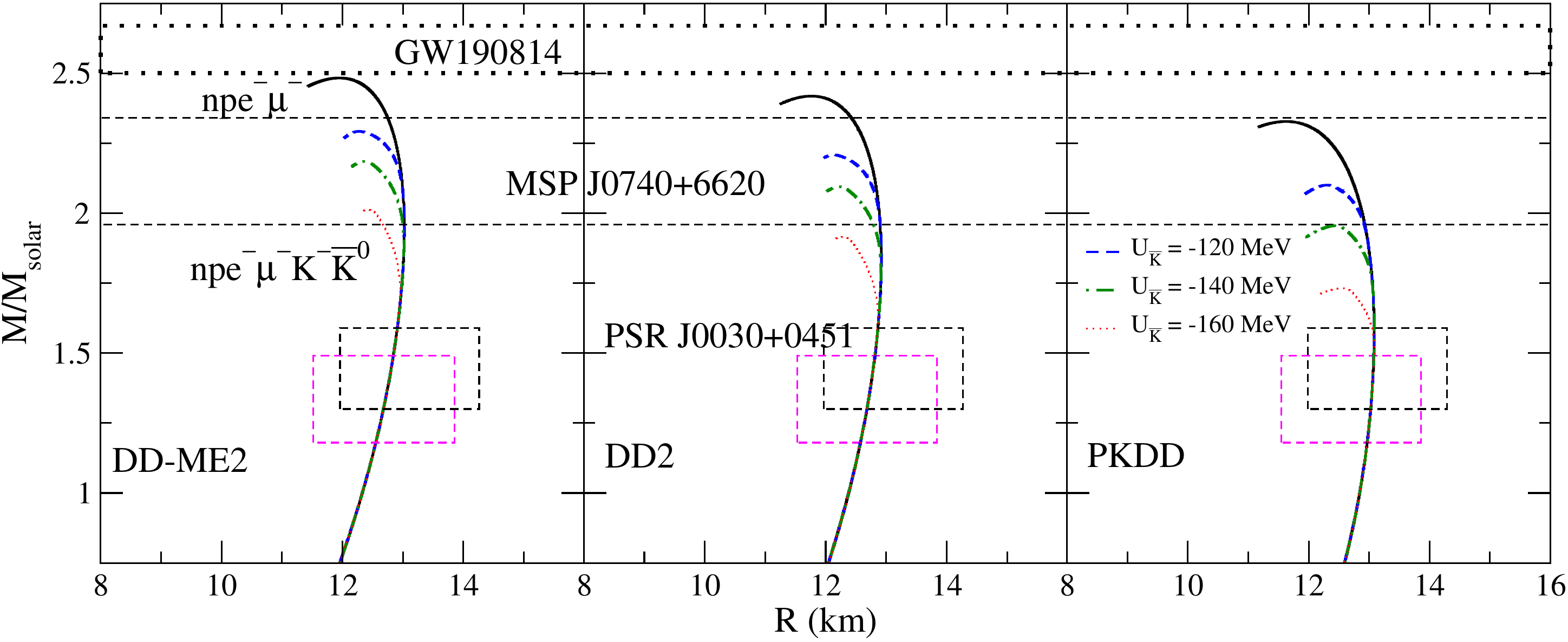}
\caption{The M-R relations corresponding to the EOSs in fig.\ref{fig-15}. Left panel: DD-ME2, center panel: DD2, right panel: PKDD parametrization. The mass-radius constraints are similar as in fig.\ref{fig-12}. The solid curve denotes the case with only nucleons and leptons, while the dashed curves correspond to $U_{\bar{K}}=-120$ MeV, dash-dotted curves represents $U_{\bar{K}}=-140$ MeV and dotted curves denote $U_{\bar{K}}=-160$ MeV case.}
\label{fig-10}
\end{center}
\end{figure*}

Fig.\ref{fig-10} presents the results of mass-radius (M-R) relationship for static spherically symmetric stars by solving the TOV equations corresponding to the EOSs in fig.\ref{fig-15}. The crust EOS is the same as considered in the NL CDF model. Table-\ref{tab:8} provides the set of maximum mass, and corresponding radius and central density with different values of $U_{\bar{K}}$. For the stars with only nucleons and leptons, the maximum masses are $2.48$, $2.42$, $2.33$ M$_{\odot}$ with DD-ME2, DD2 and PKDD parametrization respectively. The maximum mass of the compact stars decreases substantially with the inclusion of (anti)kaons. It is observed that configurations with DD-ME2 satisfy the observed maximum mas neutron star constraint of $\sim 2$ M$_{\odot}$ uptil $U_{\bar{K}}=-160$ MeV. For the other two model cases, DD2 satisfy the constraint uptil $U_{\bar{K}}=-140$ MeV, while PKDD providing a further softer EOS satisfy the constraint only upto $U_{\bar{K}}=-120$ MeV.

\begin{figure} [h!]
  \begin{center}
\includegraphics[width=8.5cm,keepaspectratio ]{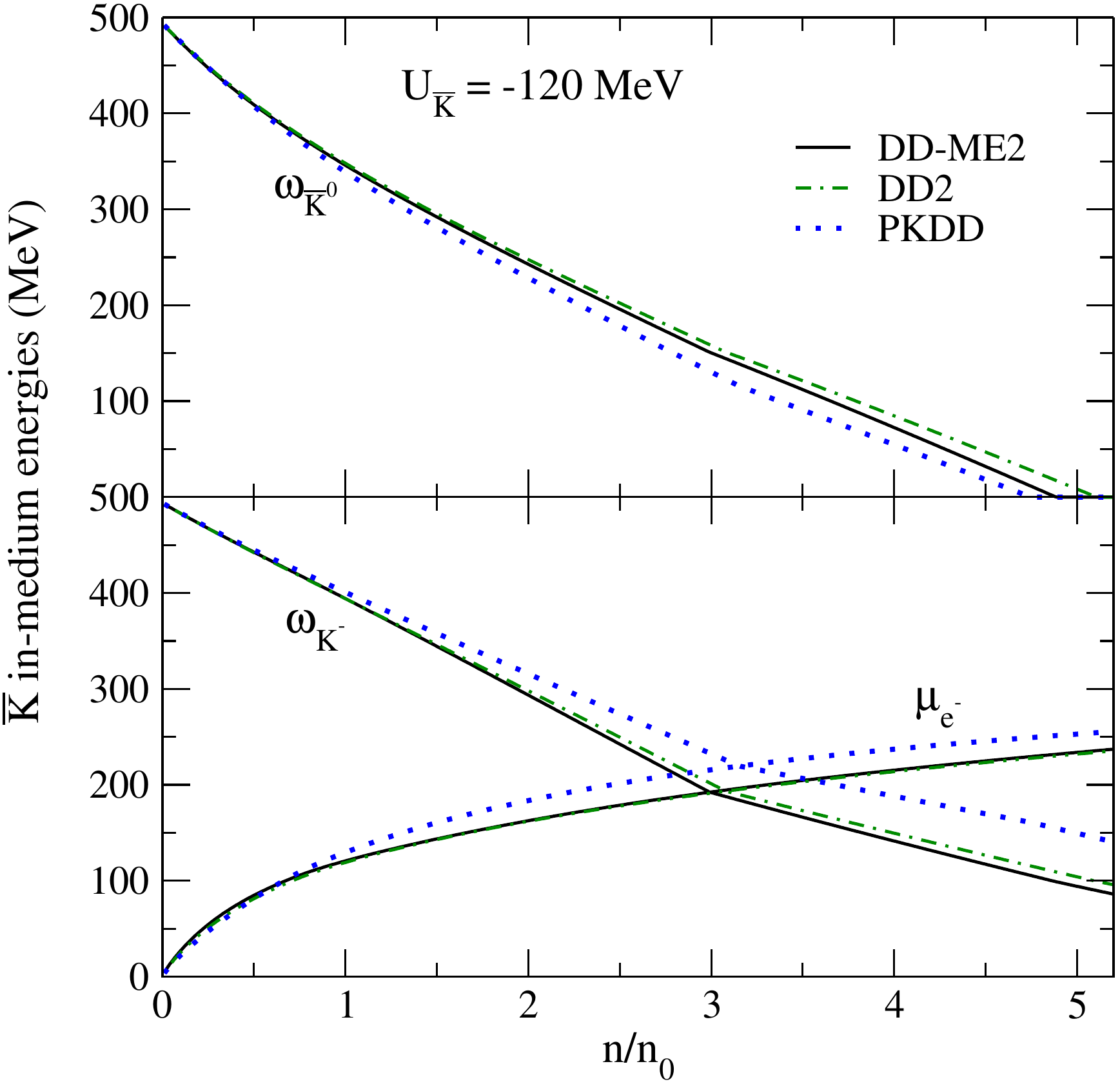}
\caption{The effective energy of (anti)kaons as a function of baryon number density with $U_{\bar{K}}=-120$ MeV. Upper panel: for $\bar{K}^0$ and lower panel: for $K^-$ with solid lines denoting DD-ME2, dash-dotted curves representing DD2 and dotted lines representing PKDD parametrization.}
\label{fig-16}
\end{center}
\end{figure}

Fig.\ref{fig-16} represents the (anti)kaon energies as a function of baryon number density with $U_{\bar{K}}=-120$ MeV for the density-dependent coupling models. The onset of $K^-$ and $\bar{K}^0$ occurs with $\omega_{K^-}$ crossing over $\mu_{e^-}$ and $\omega_{\bar{K}^0}$ being equal to zero respectively. Similar behavior is observed in case of DD-ME2 and DD2 parametrization models, while for PKDD model, the $K^-$ in-medium energy is higher compared to the other two cases and $\bar{K}^0$ in-medium energy is observed to be lower.

\begin{table*} [t!]
\centering
\caption{Maximum mass, M$_{max}$ (in units of M$_\odot$), radius (in km), corresponding central density (in units of $n_0$) of nucleon compact stars for different values of $K^-$ optical potential depths $U_{\bar{K}}$ (in units of MeV) at $n_0$ in DD-ME2, DD2, PKDD parametrization models.}
\begin{tabular}{c|c|ccc|ccc|ccc}
\hline \hline
 & & \multicolumn{3}{c}{DD-ME2} & \multicolumn{3}{|c}{DD2} & \multicolumn{3}{|c}{PKDD} \\
 \cline{3-11}
Configuration & $U_{\bar{K}}$ (MeV) & M$_{max}$(M$_\odot$) & R (km) & $n_c$ ($n_0$) & M$_{max}$(M$_\odot$) & R (km) & $n_c$ ($n_0$) & M$_{max}$(M$_\odot$) & R (km) & $n_c$ ($n_0$) \\
\hline
$npe\mu$ & 0 & 2.48 & 11.96 & 5.36 & 2.42 & 11.77 & 5.69 & 2.33 & 11.63 & 5.94 \\
\hline
 & $-120$ & 2.29 & 12.28 & 5.37 & 2.21 & 12.14 & 5.68 & 2.10 & 12.31 & 5.54 \\
$npe\mu \bar{K}$ & $-140$ & 2.18 & 12.37 & 5.33 & 2.09 & 12.23 & 5.62 & 1.95 & 12.44 & 5.36 \\
 & $-160$ & 2.01 & 12.43 & 5.22 & 1.92 & 12.29 & 5.47 & 1.73 & 12.56 & 5.05 \\
\hline
\end{tabular}
\label{tab:8}
\end{table*}

\begin{figure} [h!]
  \begin{center}
\includegraphics[width=8.4cm,keepaspectratio ]{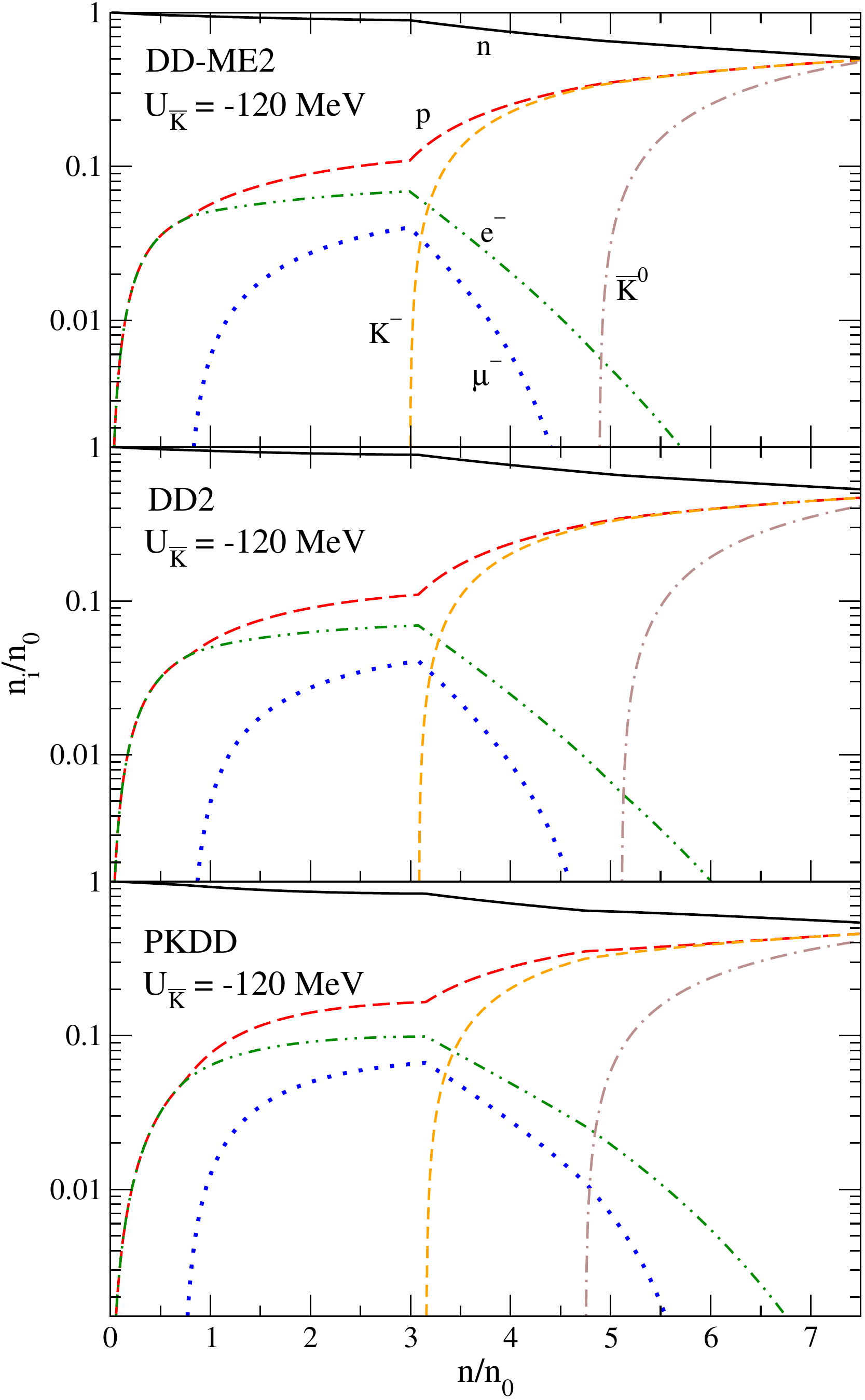}
\caption{Same as fig.\ref{fig-1}. Upper panel: DD-ME2, center panel: DD2, lower panel: PKDD parametrization model and $K^-$ potential depth of $-120$ MeV. Solid curves denote neutron ($n$), long-dashed curves proton ($p$), dash-dotted lines electron ($e^-$), dotted lines muon ($\mu^-$), short-dashed curves $K^-$, dash-double dotted lines denote $\bar{K}^0$ population.}
\label{fig-17}
\end{center}
\end{figure}

The population densities of different species, $n_i$ (in units of $n_0$) in the neutron star interior as a function of baryon number density for the density-dependent models are provided in fig.\ref{fig-17}. (Anti)kaons, being bosons are not constrained by Pauli blocking, resulting in lepton fraction suppression at high density regions. The population behavior in two cases of DD-ME2 and DD2 are observed to be similar with slight difference in the threshold density of $\bar{K}$-meson appearance. The proton population (subsequently electron and muon) is higher at lower densities for PKDD model case compared to DD-ME2, DD2 models due to higher symmetry energy. For PKDD model, in addition to the early onset of $\bar{K}^0$ particles, the eradication of lepton species is quite delayed compared to the former parametrizations. This results in further softening of EOS.


\begin{figure*} [t!]
  \begin{center}
\includegraphics[width=13.5cm,keepaspectratio ]{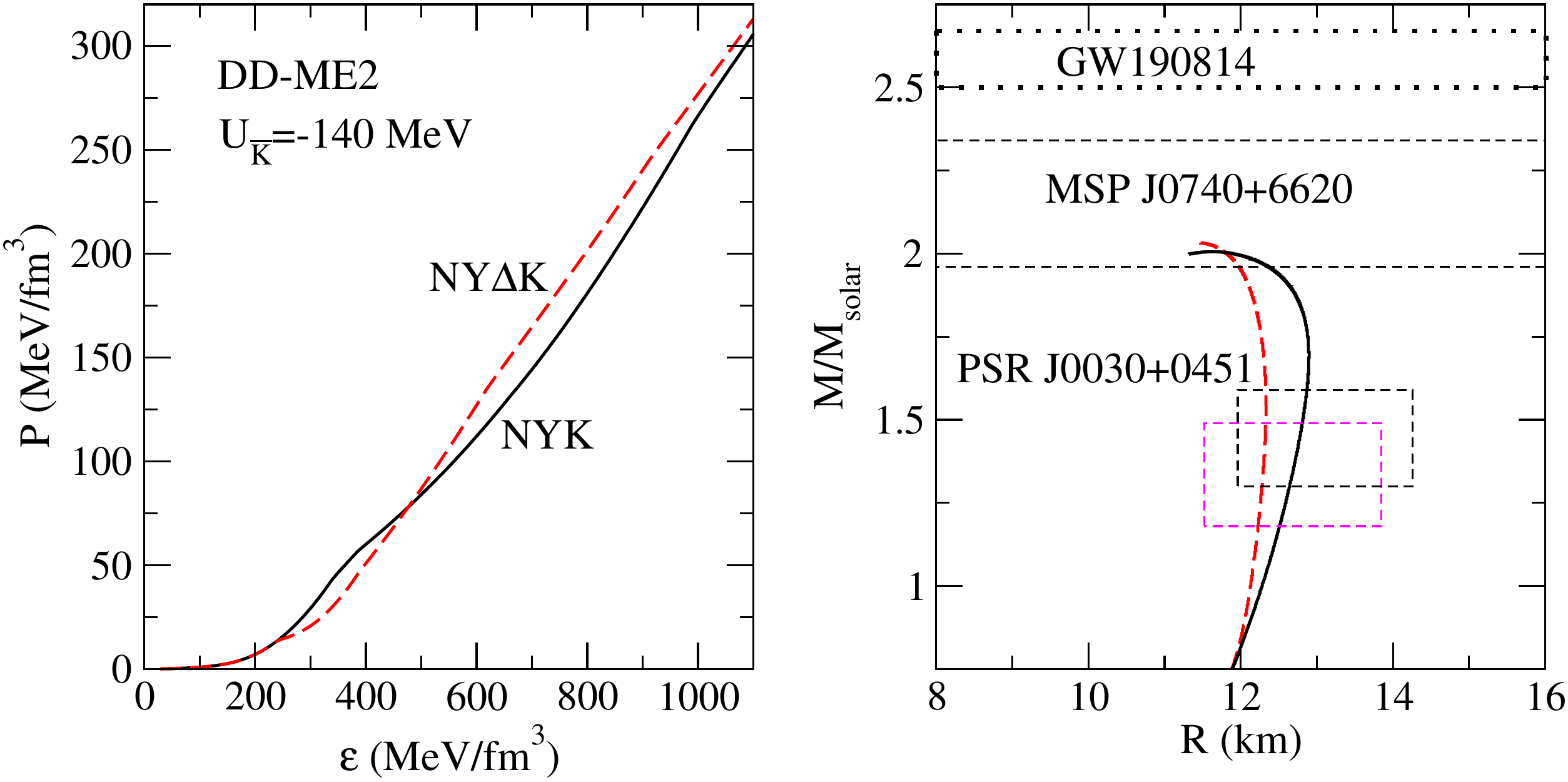}
\caption{Left panel: Equation of states, right panel: M-R relations of the neutron star matter with composition as, Nucleons, Hyperons, (anti)-kaons (NYK) and Nucleons, Hyperons, $\Delta$-resonances, (anti)-kaons (NY$\Delta$K) represented by solid and dashed lines respectively within DD-ME2 model. The mass-radius constraints are similar as in fig.\ref{fig-12}. The (anti)-kaon optical potential is considered to be $-140$ MeV.}
\label{fig-19}
\end{center}
\end{figure*}

The inclusion of hyperons as well as $\Delta$-resonances alongwith (anti)-kaons in neutron star matter in case of DD-ME2 coupling parametrization model is shown in fig.\ref{fig-19}. In this case, the meson-baryon as well as the meson-(anti-kaon) interactions are considered to be mediated by $\sigma, \omega, \rho, \sigma^*, \phi$-mesons. The coupling parameters for the meson-hyperon and meson-$\Delta$ baryon interactions are adapted from reference-\cite{particles3040043}. The EOS with matter composition as nucleons, hyperons, $\Delta$-resonances, (anti)-kaons (NY$\Delta$K) is observed to be stiffer than the one without $\Delta$-baryons (NYK). This happens because of the early appearance of $\Delta$-resonance particles which pushes the advent of hyperons to higher densities. The phase transition of (anti)-kaons is observed to be second-order form in this matter composition indicating the absence of mixed phase. Detailed discussions on this matter composition will be provided in future works.

\section{Conclusions and Outlook} \label{sec:conclusions}

We investigate the appearance of the (anti)kaon condensation in $\beta$-equilibrated charge neutral nucleonic matter within the framework of mean field theory with density-independent (NL CDF model) as well as dependent (DDRH model) coupling constants. Observations constrain the maximum mass of neutron star family to be above $2$ M$_\odot$. Stars with mass near and above $2$ M$_\odot$ must contain central density above $4~n_0$. At such high density the phase transition to boson condensate within the nuclear matter is highly probable. However, the appearance of (anti)kaon condensation softens the equation of state lowering the maximum mass of the neutron star family. We discuss here certain parametrizations of EOS within RMF model which are stiff enough to provide maximum mass above $2$ M$_\odot$ with appearance of (anti)kaon condensation at the inner core of the star. We obtain the result that with NL model, GMT and NL3 parametrizations provide the required stiff EOS to have maximum mass $2$ M$_\odot$ with all the kaon optical potential depth $U_{\bar{K}}=-120,-140,-160$ MeV. The observational constraint from GW190814 is uncertain as the star has not been identified as NS unambiguously \cite{2020ApJ...896L..44A, 2020PhRvD.102d1301S, 2020arXiv201002901J}. The (anti)kaon condensed star with $U_{\bar{K}}=-120$ MeV and NL3 coupling model persuades this gravitational wave observation mass constraint. For DDRH model, the DD-ME2 parametrization gives the observationally possible M-R relation for all the considered values of $U_{\bar{K}}$.
In most of the cases, the transition to condensed phase is through second order phase transition. First order phase transition occur only with $U_K=-160$ MeV for NL model with GMT parametrization. In this case the star possesses a mixed phase region containing both hadronic and condense phase simultaneously. To explain the mixed phase regime, Gibbs rules alongside global baryon number conservation and charge neutrality are exploited. The mixed phase region ranges for a length of $\sim 1.21$ km from stellar radius of 6.13 to 7.34 km. The outer core region upto 7.34 km is the pure hadronic phase while the region in the inner core from 6.13 km to center of the star is the pure (anti)kaon condensed phase. Moreover, the $\bar{K}^{0}$ condensation is a second-order phase transition. The compact star with the mixed phase regime satisfies the bounds set on mass-radius by the various recent astrophysical observations. The EOS (mixed phase) evaluated with GMT model incorporating the $\bar{K}$ condensation which satisfies $2$ M$_\odot$ criteria can be employed to study the glitch phenomena in pulsars. 

In case of density-dependent parametrization models (DD-ME2, DD2, PKDD), the (anti)kaon condensation is through second-order phase transition. Among these parametrizations, DD-ME2 produces the stiffest equation of state for both the cases with only nucleons as well as nucleons and (anti)kaons. All parameter sets explain the $2$ M$_{\odot}$ neutron star without the inclusion of (anti)kaons. Higher optical potential leads to early appearance of $K^-$ in the star interior. In case of DD2 model, the configuration with $U_{\bar{K}}=-160$ MeV doesn't satisfy the $\sim 2$ M$_{\odot}$ maximum mass star. For the PKDD model producing the softest EOS among the considered coupling models, the astrophysical maximum mass constraint is not satisfied with $U_{\bar{K}} \geq -140$ MeV. Within the framework model considered in this work, the upper limit for $U_{\bar{K}}$ in case of DD-ME2 model is $-160$ MeV. The likelihood of hyperon and $\Delta$-baryons in the neutron star matter alongside (anti)-kaons are also studied with DD-ME2 model and $U_{\bar{K}}=-140$ MeV. The (anti)-kaons tend to appear at the higher density for the matter with $\Delta$-baryons in comparison to the one without $\Delta$-resonances in the neutron star matter. The mass-radius constraints are observed to be satisfied by this matter composition as well. Further analysis on the effects due to the presence of hyperons as well as $\Delta$-baryons with the appearance of (anti)-kaonic condensation in neutron star matter is beyond the scope of this work.

\begin{acknowledgements} 
The authors thank the anonymous referee for the constructive comments which have contributed to ameliorate the quality of the manuscript significantly. The authors acknowledge the funding support from Science and Engineering Research Board, Department of Science and Technology, Government of India through Project No. EMR/2016/006577 and Ministry of Education, Government of India. The authors are also thankful to Sarmistha Banik and Debades Bandyopadhyay for fruitful discussions. 
\end{acknowledgements}

\bibliography{draft}
\end{document}